\documentclass[%
 reprint,
 table,
superscriptaddress,
nofootinbib,
 amsmath,
 amssymb,
 aps,
 prapplied,
floatfix,
]{revtex4-2}

\usepackage{graphicx}
\usepackage{lipsum}
\usepackage{hhline}
\usepackage{booktabs}
\usepackage{multirow}
\makeatletter
\newcommand*\bigcdot{\mathpalette\bigcdot@{.5}}
\newcommand*\bigcdot@[2]{\mathbin{\vcenter{\hbox{\scalebox{#2}{$\m@th#1\bullet$}}}}}
\makeatother

\usepackage{dcolumn}
\usepackage{bm}

\usepackage[T1]{fontenc}
\usepackage{amssymb}
\usepackage{mathtools}
\usepackage{MnSymbol}
\usepackage{bbold}
\usepackage[position=top, caption=false]{subfig}
\usepackage[table,x11names,dvipsnames,table]{xcolor}
\usepackage{tikz}
\usetikzlibrary {shapes.geometric}

\usepackage{amsthm}
\usepackage{amsmath}
\usepackage{mathrsfs}
\usepackage{dsfont}
\usepackage{physics}
\usepackage[thinc]{esdiff}
\usepackage{nicefrac}
\usepackage{xfrac}
\usepackage{algorithm}
\usepackage{algpseudocode}
\usepackage[pdftex, pdftitle={Article}, pdfauthor={Author}]{hyperref}
\usepackage[capitalise]{cleveref}
\creflabelformat{equation}{#2\textup{#1}#3}

\begin{document}
\preprint{APS/123-QED}

\title{Almost fault--tolerant quantum machine learning with drastic overhead reduction}

\author{Haiyue Kang}
 \email{haiyuek@student.unimelb.edu.au}
 \affiliation{School of Physics, University of Melbourne, VIC, Parkville, 3010, Australia}
 
\author{Younghun Kim}
 \affiliation{School of Physics, University of Melbourne, VIC, Parkville, 3010, Australia}

\author{Eromanga Adermann}
 \affiliation{Quantum Systems, Data61, CSIRO, Australia}
 
\author{Martin Sevior}
 \affiliation{School of Physics, University of Melbourne, VIC, Parkville, 3010, Australia}

\author{Muhammad Usman}
 \affiliation{Quantum Systems, Data61, CSIRO, Australia}
 \affiliation{School of Physics, University of Melbourne, VIC, Parkville, 3010, Australia}

\begin{abstract}
Errors in the current generation of quantum processors pose a significant challenge towards practical-scale implementations of quantum machine learning (QML) as they lead to trainability issues arising from noise-induced barren plateaus, as well as performance degradations due to the noise accumulation in deep circuits even when QML models are free from barren plateaus. Quantum error correction (QEC) protocols are being developed to overcome hardware noise, but their extremely high spacetime overheads, mainly due to magic state distillation, make them infeasible for near-term practical implementation. This work proposes the idea of partial quantum error correction (QEC) for quantum machine learning (QML) models and identifies a sweet spot where distillations are omitted to significantly reduce overhead. By assuming error-corrected two-qubit Controlled-$Z$s (Clifford operations), we demonstrate that the QML models remain trainable even when single-qubit gates are subjected to $\approx0.2\%$ depolarizing noise, corresponding to a gate error rate of $\approx0.13\%$ under randomized benchmarking. Further analysis based on various noise models, such as phase-damping and thermal-dissipation channels at low temperature, indicates that the QML models are trainable independent of the mean angle of over-rotation, or can even be improved by thermal damping that purifies a quantum state away from depolarizations. While it may take several years to build quantum processors capable of fully fault-tolerant QML, our work proposes a resource-efficient solution for trainable and high-accuracy QML implementations in noisy environments.

\end{abstract}

\maketitle
\section{Introduction}\label{sec:introduction}
Inspired by classical artificial intelligence, quantum machine learning (QML) \cite{QML,QML2,VQA,VQE,QAOA,QPCA,QNN} is a topic of highly active research, as it aims to explore the utility at the intersection of both quantum computing and the ideas of machine learning. Advances in the field of QML have been in fundamental algorithmic development \cite{QML_full_summary,QML,QML2,VQA,VQE,Accelerated_VQE,VQE_variants_summary,VQSE,QAOA,QPCA,QNN,QNN_2,VQC,QVC,QVC_2,QGAN,QGAN2,TN_QML,quantum_chemistry,QCNN_classically_simulable,reflection_invariant_qml,rotational_invariant_qml,non_unitary_qml,Zehang_Wang_QML_1,barren_plateaus} or in small-scale empirical demonstrations using idealized (noise-free) quantum simulators \cite{VQSE,Zehang_Wang_QML_1,Max_West_adv_QML_1,Max_West_adv_QML_2,QNN_2,VQC,TN_QML,reflection_invariant_qml,rotational_invariant_qml,non_unitary_qml,barren_plateaus,emprical_VQE,thermal_helps_learning,thermal_helps_learning2}. Despite the availability of a wide range of quantum devices, the implementation of QML on quantum hardware is scarce \cite{QML_full_summary,johri2021nearest,wang2021towards} which is mainly due to the high level of noise or errors in the current generation of quantum computers and very deep circuits associated with the QML architectures -- both factors being highly detrimental to the QML performance. As for every other quantum algorithm, the key to achieving a quantum advantage in QML is to enable its fault-tolerant implementation, which will rely on quantum error correction codes to overcome hardware noise. The full-scale quantum error correction (QEC) for QML will be highly resource intensive \cite{surface_codes,qec_lattice_surgery}, requiring millions of qubits, and therefore years if not decades away from practical realization. This work presents a novel scheme to implement QML models with nearly fault-tolerant accuracies on noisy quantum processors. Our proposal will drastically reduce QEC resource requirements at the expense of a small reduction in QML accuracy, bringing practical QML much closer to reality. 

\begin{figure*}[thbp!]
    \centering
    \subfloat[\label{fig:qml_scalability}Behavior of QNN in various scenario]{\includegraphics[width=0.45\linewidth]{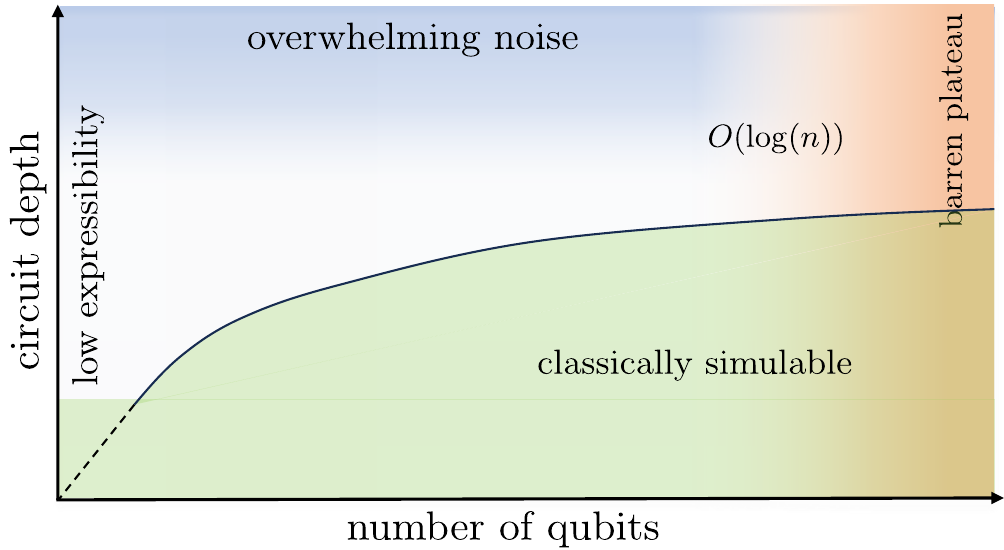}}\\
    \subfloat[\label{fig: Normal loss function landscape}Normal loss function]{\includegraphics[width=.333\linewidth]{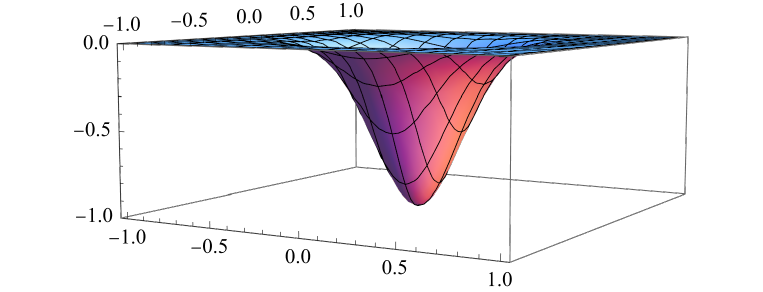}}
    \subfloat[\label{fig: Barren plateaus}Barren plateaus]{\includegraphics[width=.333\linewidth]{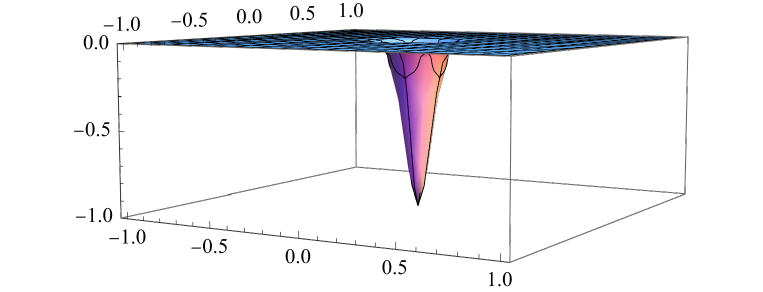}}
    \subfloat[\label{fig: Noise-induced barren plateaus}Noise-induced barren plateaus]{\includegraphics[width=.333\linewidth]{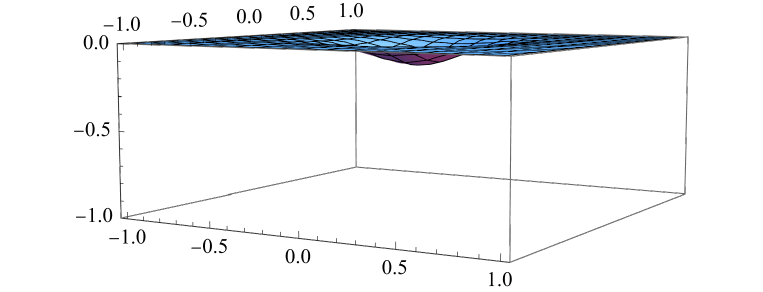}}
    \caption{(a) An illustration of the major obstacles of a quantum variational classifier on different scales of circuit depth and number of qubits: The model becomes non-trainable due to the scaling in the number of qubits, as denoted by barren plateaus in the orange-shaded region. When the circuit depth scales faster than $O(\log(n))$, the quantum circuit is easily overwhelmed by the noise, leading to noise-induced barren plateaus as denoted by the blue-shaded region. In contrast, the circuit can be efficiently simulated classically if the circuit depth scales more slowly than $O(\log(n))$. If only a few qubits are involved, it also lacks expressibility for encoding the state to demonstrate the utility of quantum computation. (b) Example of a well-behaved landscape of a loss function for a trainable quantum neural network, where the function values and parameter values are represented on the vertical and horizontal axes, respectively. (c) The typical landscape for conventional barren plateaus without any noise, where the gradients are vanishing almost everywhere except some small regions of well-behaved parameters, typically around the minima \cite{BP_narrow_gorges,barren_plateaus,barren_plateaus_summary}. (d) The landscape for noise-induced barren plateaus, where there are no exceptions to gradients that are non-vanishing. }
    \label{fig:landscapes_comparison}
\end{figure*}

In recent years, a variety of QML models have been developed and characterized for various datasets, with the most widely adopted model being the quantum variational classifier (QVC) \cite{QVC,QVC_2,VQC}, known as parametrized quantum circuits (PQC). These models are constructed from multiple layers of single-qubit parametrized or rotation gates, which are tuned during the QML training process, and unparametrized two-qubit gates, which introduce entanglement to the circuit. The basic working principle of QVC-based QML is quite similar to classical machine learning in the sense that a loss function is iteratively minimized, driving towards its global minimum, which is usually achieved in the form of fitting some highly non-trivial parametrized function. These QVC architectures, despite demonstrating promising learning abilities at the proof-of-concept level, suffer from several challenges. One major issue is related to the trainability of these models, which arises from barren plateaus \cite{barren_plateaus, barren_plateaus_summary} that lead to loss function gradients concentrated near zero or exponentially vanishing as the number of qubits increases. In addition, scaling in the circuit depth beyond the $O(\log(n))$ layers also leads to quantum circuits being overwhelmed by noise, resulting in gradients too small to be distinguishable from statistical fluctuations. This is known as noise-induced barren plateaus \cite{noise_induced_barren_plateaus}. In contrast, a model with a small number of qubits and low circuit depth would be easily classically simulable \cite{QCNN_classically_simulable, classical_simulation}, or it would have low expressibility and lose potential quantum utility. Figure \ref{fig:qml_scalability} shows a visual illustration of these challenges.

Although the literature has shown that the conditions for barren plateaus could be relaxed to higher limits with careful treatment \cite{local_cost_function_1, local_cost_function_2, pre_training_avoids_barren_plateaus, layer_by_layer_1, layer_by_layer_2}, and there are provably trainable models \cite{reflection_invariant_qml,rotational_invariant_qml}. However, such models have not been tested in the presence of hardware noise, especially for deep circuits \cite{shallow_VQC_noise_mitigation}. In fact, these protocols remain vulnerable to barren plateaus induced by noise \cite{noise_induced_barren_plateaus, local_cost_function_1}, where the landscapes of the loss function are suppressed everywhere. This is summarized in \Cref{fig: Noise-induced barren plateaus}. These plateaus prevent training from finding the global minimum. Therefore, regardless of whether barren plateaus are resolved in a noise-free simulation, it is essential to consider the problem under noisy conditions.

\begin{table*}[t!]
    \centering
    \resizebox{\textwidth}{!}{
    \begin{tabular}{c|c|c|c|c|c|c|c|c|c|c}
         \toprule 
         \multirow{2}{*}{\shortstack{Error\\budget}}
         & \multirow{2}{*}{\shortstack{QVC\\layers}}
         & \multirow{2}{*}{\shortstack{logical qubit\\error rate\\(per cycle)}}
         & \multirow{2}{*}{\shortstack{distilled $T$ \\error rate}}
         & \multicolumn{4}{c}{Spatial Cost}\vline & \multicolumn{3}{c}{Temporal Cost}
         \\\cline{5-11}
         &&&&\shortstack{Code\\distance}&\shortstack{Data\\qubits} & \shortstack{$T$ factory\\qubits}& \shortstack{$T$ factory qubits\\(no distillation)} & \shortstack{One logical\\cycle (\text{$\mu$}s)} & \shortstack{No. logical\\cycles} & \shortstack{Total\\runtime (ms)}
         \\\hline
        
         $1\times10^{-3}$ &50&$3.00\times10^{-10}$&$2.47\times10^{-9}$&15 &13500 & $1.746\times10^6$ & $\sim1.35\times 10^4$&6.0&4060&24\\\cline{1-11}
         $1\times10^{-3}$&100&$3.00\times10^{-10}$&$2.47\times10^{-9}$&15  &13500 & $1.782\times10^6$ & $\sim1.35\times 10^4$&6.0&8410&50\\\cline{1-11}
         $1\times10^{-4}$&50&$3.00\times10^{-11}$&$5.51\times10^{-10}$&17  &17340 & $1.746\times10^6$ & $\sim1.73\times 10^4$&6.8&4360&30\\\cline{1-11}
         $1\times10^{-4}$&100&$3.00\times10^{-11}$&$5.51\times10^{-10}$&17  &17340 & $1.764\times10^6$ & $\sim1.73\times 10^4$&6.8&8710&59         
         \\\bottomrule
    \end{tabular}
    }
    \caption{Summary on the cost of the full variational circuit with 10 algebraic qubits ($Q_{\text{alg}}=10$) with different numbers of layers and logical qubits with or without distillation, evaluated by the Azure quantum resource estimator \cite{Azure_quantum_resource_estimator, Azure_quantum_resource_estimator_paper}. For the situation that involves distillations, the figures are calculated based on a physical gate error rate of $10^{-3}$ for both the Clifford gates and $T$ gates before distillation, while achieving a specific logical error rate for the whole variational circuit. For the situation without distillation, the cost of the $T$ factory qubits is simply a number at the same level as the number of data qubits, since they share the same code distance, and there are a maximum of $Q_{\text{alg}}$ logical $\ket{T}$ states to be injected in parallel. It is easy to see that the total overhead differs by almost 2 orders of magnitude between with and without distillation. To a large extent, the time cost is linear with respect to the number of QVC layers. This should still be the case without distillation, yet with a significantly smaller time cost per logical patch-cycle.}
    \label{tab: full VQC costs}
\end{table*}
\begin{figure*}[t!]
    \centering
    \includegraphics[width=\linewidth]{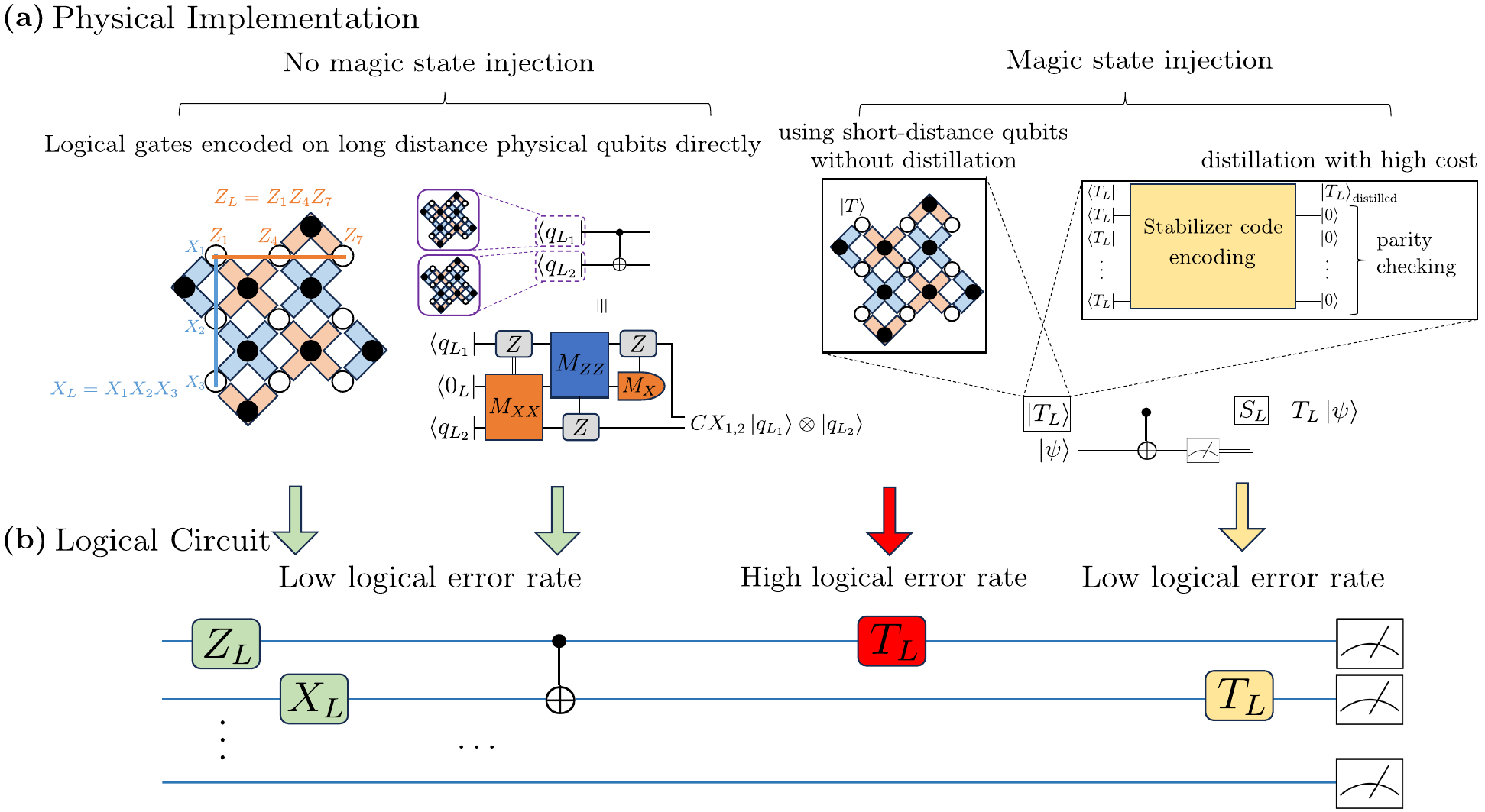}
    \caption{(a) Demonstration of how logical operators are implemented on patch-based surface codes introduced in Ref. \cite{surface_codes}. White dots indicate data qubits, black dots indicate syndrome extraction qubits with orange and blue strips representing Pauli-$X$ and $Z$ stabilizers, respectively. For operators induced from the Clifford group, including $X$, $Z$ and $CX$ gates, their logical operators can be encoded from many physical operators directly, without the need for ancilla qubits \cite{surface_codes, qec_lattice_surgery}. For operators outside the Clifford group, such as the $T$ gate, its logical operator $T_L$ cannot be implemented directly, but must be teleported from an ancilla logical qubit in the state $\ket{T_L}=\frac{1}{\sqrt{2}}(\ket{0_L}+e^{i\pi/4}\ket{1_L})$ via magic state injection. It turns out that $\ket{T_L}$ must be prepared from a single, physical qubit $\ket{T}$ state first, and then perform stabilizer measurements \cite{qec_lattice_surgery}. 
    Or, one could choose to carry out magic state distillation with very high spacetime cost. (b) Without the redundancy of encoding one logical operator from multiple physical operators, the logical error rate for $T$ gates is comparable to the physical $T$ error rate. However, if the state is distilled properly, the logical gate error rate can be suppressed to the same level as other Clifford gates.}
    \label{fig: QEC_cost}
\end{figure*}

The implementation of QML with QEC can address challenges associated with trainability and noise accumulation in deep quantum circuits. However, existing QEC protocols, such as surface codes, are too expensive in terms of spacetime cost for QML architectures. Variational QML circuits rely on parametrized quantum gates, which are non-Clifford operations and typically require highly expensive logical $T$-gate implementation through magic state injection, which is based on the preparation of a logical magic state from a single physical qubit or short-distance qubit first \cite{surface_codes}. Without the redundancy of multiple physical qubits, the logical $T$-gate error rate is close to the physical qubit error rate. Thus, one must mitigate the magic state through state distillation \cite{surface_codes, magic_state_distillation_cost1, magic_state_distillation_cost2}, yet this leads to an exponentially large spacetime overhead. Through a careful analysis, we find that to simply execute a 10-logical qubit QVC circuit, millions of physical qubits are required in the $T$-state factory with code distance greater than 15 to achieve a reasonable full-circuit logical error rate, which is two orders of magnitude larger than without distillation. Even after optimizations in some of the recent distillation protocols \cite{magic_state_distillation_cost1, magic_state_distillation_cost2}, the cost is on the order of 10000 qubit rounds for every instance of distillation, accumulating to millions of qubits for the whole QEC process. In addition to magic state distillation, one may also consider the technique of magic state cultivation \cite{magic_state_cultivation}. The success rate of preparing a logical magic state with a low error rate is extremely low, and so its actual space-time cost is still much higher than injecting a raw magic state. Consequently, fault-tolerant QML with the full-scale QEC protocol is nonviable on near- to medium-term quantum processors.

To overcome the challenges stemming from barren plateaus, hardware noise, and implementation on near-term quantum processors, we take a step back from a fully fault-tolerant surface code-based QEC and propose a novel approach that relies on partial QEC. In our scheme, only Clifford gates (including Controlled-$Z$s) are error-corrected, and the single-qubit parametrized gates are left without any noise cancellation. This eliminates the major cost of QEC associated with non-Clifford operations, thereby reducing resource requirements from millions of qubits to merely a few thousand qubits. Our scheme is motivated by the fact that single-qubit parametrized gates are being trained, and therefore, any noise-induced disturbances will be overcome by the training process itself. This is partially inspired by classical machine learning studies, where it has recently been shown that neural networks can be trained regardless of noisy neurons \cite{noisy_classical_CNN,stochastic_classical_NN}. Indeed, our work demonstrates that QVC models can be trained with error-corrected Controlled-$Z$s and noisy single-qubit gates without encountering any trainability issues and without compromising final classification accuracy.  

To demonstrate the operation of our scheme, we selected the 10-class MNIST dataset \cite{MNIST}, which is a standard benchmark in the literature for classical and quantum machine learning models. We experimentally demonstrate the trainability of the QVC75 model (a model with 75 variational layers) for a single-qubit gate error rate comparable to that of state-of-the-art quantum computers, which is reinforced by gradient values above the shot noise. In the simulation, the performance of the model is only slightly degraded compared to the full fault-tolerant case, highlighting that our model is almost fault-tolerant. Additionally, we demonstrate that a drastic reduction in spacetime overhead can be achieved without requiring distillation, as summarized in \Cref{tab: full VQC costs,tab: distillation costs}. Our work can be generalized to deeper QVC models to further improve classification accuracies, as well as to novel architectures with provable trainability in the presence of hardware noise. On the computational side, QML simulations with noise are highly expensive tasks that require significant computational resources and time. In our study, QVC training for each noise configuration took approximately 3-6 days of GPU time. This might be one of the primary reasons for the lack of such investigation in the literature. The key novelty of our work is in the discovery that QML models are naturally resilient to single-qubit noise and therefore do not require full QEC, allowing us to save significant overhead. This is not true for any other quantum algorithm and has not been shown previously in our knowledge. On top of that, this work is the first comprehensive study of QML subjected to a variety of noise mechanisms, providing critical new insights that will be valuable in their practical application in the NISQ era and early fault-tolerant implementations. 

In the remaining sections, we explain the general framework of this paper, including the necessity of implementing partial QEC, the setup of the noisy variational circuits in \Cref{sec:framework}, and our numerically simulated results in \Cref{sec: Numerical results}. We summarize our results in \Cref{sec: Discussion}.

\section{Framework}\label{sec:framework}
\begin{figure*}[t!]
    \centering
    \subfloat[\label{fig: QNN workflow} Workflow of the quantum neural network for data classification]{\includegraphics[width=.8\linewidth]{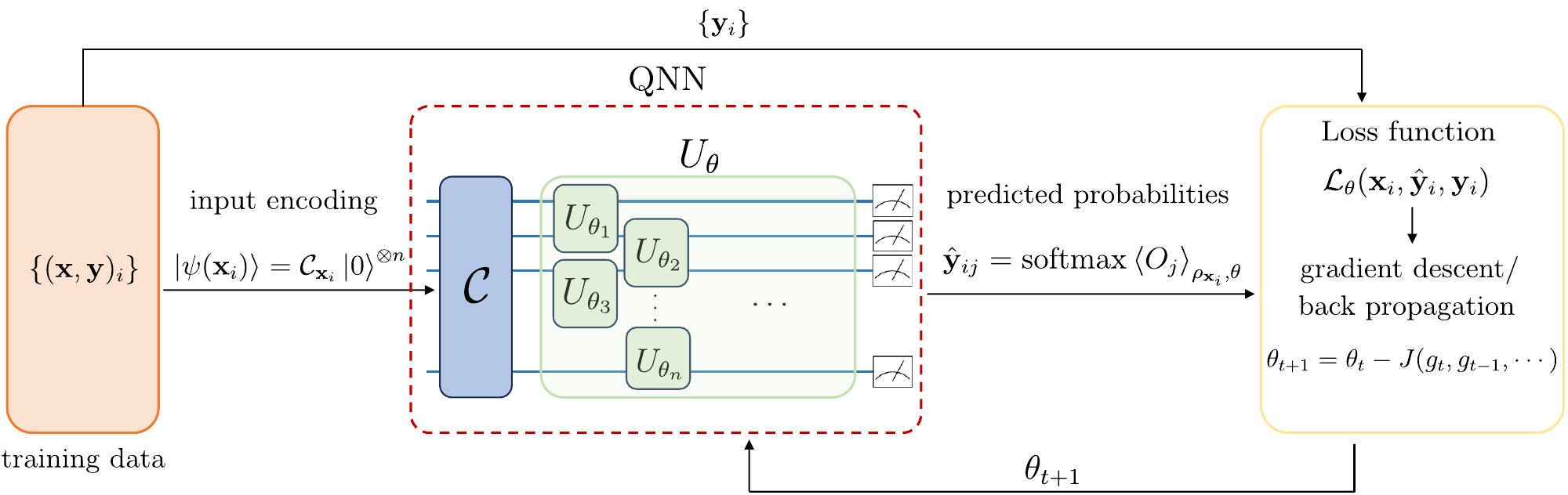}}\\
    \subfloat[\label{fig: VQC} Variational Circuit]{\includegraphics[width=.6\linewidth]{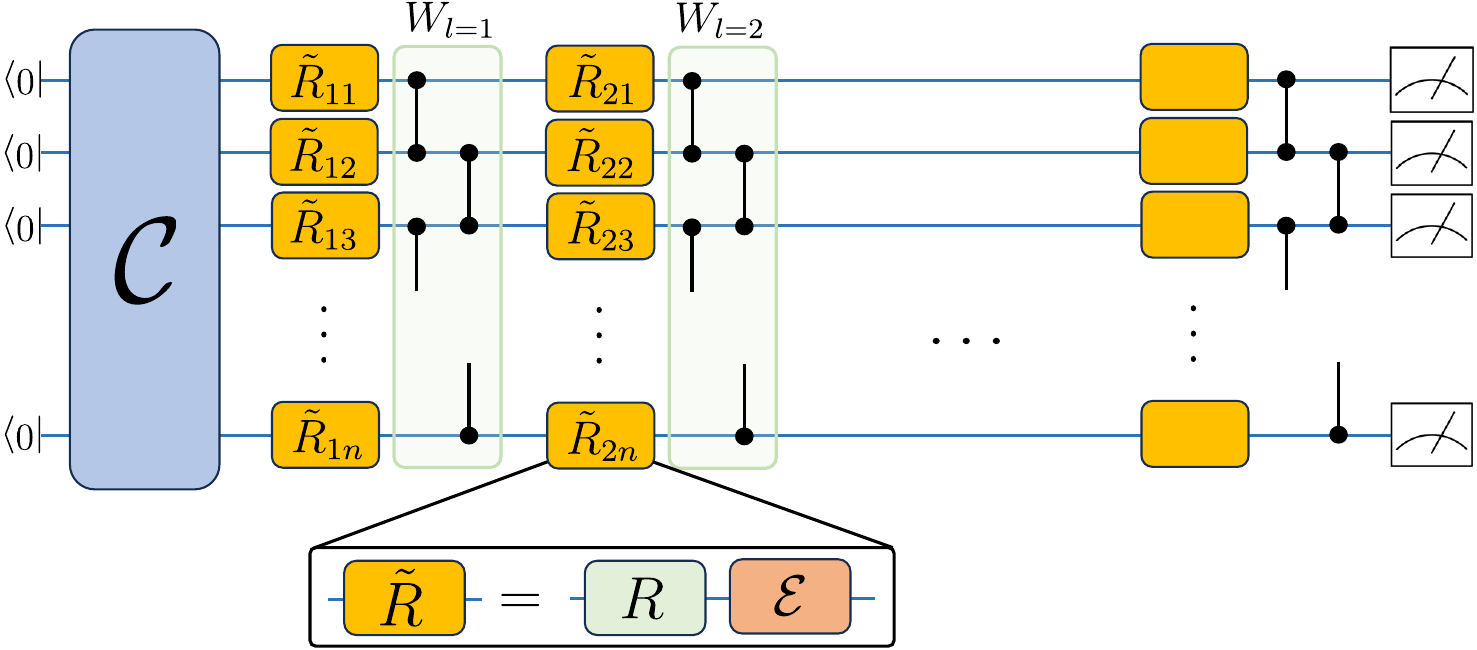}}
    \caption{(a) General workflow of a quantum neural network for classification tasks. 1. One first prepares a set of data-label pairs for training, after which 2. the data (e.g. an array of pixel values of an image) is encoded onto a quantum state $\ket{\psi(\bm{x}_i)}=\mathcal{C}(\bm{x}_i)\ket{0}^{\otimes n}$. The encoded state evolves through a sequence of parameterized unitaries followed by measurements on each qubit, and the observable expectation values correspond to the predicted probabilities of the respective label. Consequently, the predicted label is given by the qubit index with the highest expected value. 3. The predicted probabilities are compared with the true probabilities obtained from the actual label through the cross-entropy loss function, which quantitatively evaluates the inference performance. 4. The parameters are updated iteratively using the gradient descent method. (b) The detailed circuit design of the variational circuit presented in this paper. The data is encoded into the circuit through amplitude encoding to minimize the requirement for the number of qubits and conserve memory for simulation. The parameterized unitary consists of multiple layers of unitaries, with each layer containing a sequence of single-qubit rotations with parameterized rotation axes and angles, denoted as $R_{lm}$ as shorthand for $R_{m {n}_{lm}}(\theta_{lm})$, followed by a sequence of entangling Controlled-$Z$ gates without trainable parameters. An error channel $\mathcal{E}$ is added after every ideal gate $R_{lm}$, constitutes $\Tilde {R}_{lm}$. The inferred probabilities for each potential label are evaluated by measuring the Pauli-$Z$ expected values on each of the qubits.}
    \label{fig: QNN}
\end{figure*}
\subsection{Partial QEC}\label{subsec: partial QEC}
Since depolarizing noise maps to the image of the maximally mixed state, which loses all the quantum information in the long term, its mitigation cannot be accomplished through any self-adaptive mid-circuit manipulations such as dynamical decoupling \cite{dynamical_decoupling}, Pauli twirling \cite{pauli_twirling_1}, whereas other strategies, including zero-noise extrapolation \cite{zero_noise_extrapolation}, probabilistic error cancellation \cite{pauli_twirling_2}, usually require noise parameter characterizations of substantial overhead. Therefore, the necessity of quantum error correction becomes apparent when dealing with the noise-induced barren plateaus of quantum neural network training. 

However, as explained in \Cref{fig: QEC_cost}, without magic state distillation, a logical $T$ gate error rate comparable to the physical error rate is inevitable. However, there is a significant spacetime overhead associated with magic state distillations required for logical $T$ gate operations with low error rates. To evaluate this overhead in a quantum neural network, we numerically estimate the scaling of the spacetime complexities of the full variational circuit with 10 qubits, estimated through the Azure quantum resource estimator \cite{Azure_quantum_resource_estimator, Azure_quantum_resource_estimator_paper}, which is based on Ref. \cite{surface_codes}, as shown in \Cref{tab: full VQC costs}. Specifically, for a 100-layer QVC, the cost skyrockets to $1.76\times10^{6}$ physical qubits ($6090d^2$ qubits with code distance 17) in order to achieve a reasonable whole-circuit error budget of $10^{-4}$ with a physical gate error rate of $10^{-3}$. However, without distillation, it is expected that only $\sim60d^2$ qubits are needed, which is comparable to the size of data qubits, for the $T$ factory; therefore, maximally $\sim120d^2$ qubits in total are necessary. However, the time cost should still scale linearly with respect to the number of layers in the circuit, but with a significantly smaller time cost per patch-cycle, which is the number of logical patches times the logical cycles. Overall, it is easy to see that the total overhead differs by almost two orders of magnitude with and without distillation. Even when considering some of the latest magic state distillation protocols \cite{magic_state_distillation_cost1, magic_state_distillation_cost2}, the contrast to the case without any distillation is still significant, as shown in \Cref{tab: distillation costs} (See \Cref{appendix: Spacetime cost of magic state disillation,appendix: Cost of full variational circuit} for a more detailed discussion).
Clearly, despite the reduction in the logical error rate after the $L$ layers of distillations, the spatial cost also scales exponentially, and the time cost scales linearly with respect to $L$. Depending on protocols (i.e., the number of $X$ stabilizers $m_X$, and the $n:k$ ratio, etc.), the spacetime cost of distillation also varies. However, even the one with the worst improvement in logical error still requires $11$ times more logical patch-cycles with code distance $11$; that is, 14600 qubit-rounds.

Therefore, to avoid such an overhead, we propose a partial QEC code, which simply omits the distillation process, so that the noisy magic state is injected directly, and the structure of the code is maintained. By doing this, we can treat the logical variational circuit having its two-qubit gates, as long as they were induced from the Clifford group, with a negligible error rate. We remark that this approximation is exceptionally accurate: In our setting, after implementing the partial-QEC, the logical two-qubit Clifford gate can easily achieve an error rate at the level of $10^{-10}$ with reasonable code distance and physical qubits (see \Cref{tab: full VQC costs}), whereas the net logical error rate of an arbitrary noisy single-qubit unitary, which is composed with distillation-free $T$ gates, is expected to be not significantly less than $10^{-4}$ in foreseeable future ($1.33\times10^{-3}$ specifically, see discussion in \Cref{subsec: depol channel} and \Cref{appendix: Depolarising noise strength and actual gate error analysis}). This differs by more than 6 orders of magnitude, and unsurprisingly, the noise is almost certainly dominated by the single-qubit unitaries in the setting of partial-QEC. However, significant error rates remain for a general single-qubit unitary $U_{\hat{\bm{n}}}(\theta)$ with error rate $\sim1-(1-\epsilon_T)^{\log_2(1/\epsilon)}$, where $\epsilon_T$ is the raw $T$ gate error rate and the power $\log_2(1/\epsilon)$ is the number of $T$ gates involved in achieving precision $\epsilon$ \cite{T_consumption_for_unitary_improved} of the rotation angle $\theta$.

\subsection{Variational circuits with single-qubit gate noise}\label{subsec: VQC setup}
To test our hypothesis of partial QEC, we perform handwriting number classification tasks based on MNIST datasets in the quantum neural network using the supervised learning variational circuit model with various numbers of layers and qubits, plus an additional noise model in the single-qubit gates. 

As shown in \Cref{fig: QNN workflow}, the quantum neural network aims to fit a highly non-trivial and probabilistic mapping $f: \mathcal{R}\rightarrow\mathcal{Y}$ that maps some input vector $\bm{x}_i\in \mathcal{R}$ to a label vector $y_i\in\mathcal{Y}$, $|{\mathcal{Y}}|=K$, that encodes some real world classifications, through a parametrized function $h_{\bm{\theta}}:\mathcal{R}\rightarrow\mathcal{Y}$ with trainable parameters $\bm{\theta}$ that reproduces the mapping $f$ with high success probability. In our case, this would be a pair of arrays of pixel values corresponding to the picture of a handwritten number and labels indicating the number. The parameters are trained intentionally towards the (ideally) global minima $\bm{\theta}^{*}$ of some loss function $\mathcal{L}$
\begin{equation}
\bm{\theta}^{*}\coloneqq\underset{\bm{\theta}\in\Theta}{\arg\min}\sum\limits_{i}\mathcal{L}(\bm{x}_i,\hat{\bm{y}}_{\bm{\theta}}(\bm{x}_i),\bm{y}_i),
\end{equation}
where $\{(\bm{x}_i,y_i)\}\subseteq \mathcal{R}\cross \mathcal{Y}$ are the training datasets, $\hat{\bm{y}}_{\bm{\theta}}(\bm{x}_i)=(\hat{\bm{y}}_{i1},\cdots,\hat{\bm{y}}_{iK})^{T}$ are the likelihoods of each class predicted by the model given parameters $\bm{\theta}$ such that $h_{\bm{\theta}}(\bm{x}_i)\equiv \underset{j}{\arg\min}\hat{\bm{y}}_{\bm{\theta}}(\bm{x}_i)_j$, and $\bm{y}_i=(0,\cdots,1,\cdots,0)^T=\bm{e}_{y_i}$ with all zeros except one on the index $y_i$. To reach the minima $\bm{\theta}^{*}$, we implement the ADAM variant \cite{gradient_descent_algorithms_summary} of the descent method stochastically with 50 images as one batch, where
\begin{equation}
    \bm{\theta}_{t+1}=\bm{\theta}_t-J\left(\bm{g}_t,\bm{g}_{t-1},\cdots,\bm{\theta}_t,\bm{\theta}_{t-1},\cdots\right),
\end{equation}
where $\bm{g}_t=\vec\nabla_{\bm{\theta}_t}\mathcal{L}$ and $J$ is a function of the sum of the gradients $\bm{g}_k$ on the path of descent with some weights $w_k$ depending on how many iterations ago the gradient was obtained \cite{gradient_descent_algorithms_summary},
\begin{equation}
    J\left(\bm{g}_t,\bm{g}_{t-1},\cdots,\bm{\theta}_t,\bm{\theta}_{t-1},\cdots\right)=\sum\limits_{k=1}^{t}w_k g_k.
\end{equation}
For the ADAM variant, the closer $k$ is to the current iteration $t$, the larger $g_k$ is. The only component of this algorithm that involves quantum computation occurs in the evaluation part, where the computation of $\hat{y}_{\bm{\theta}}$ is executed on a quantum circuit during the training process \cite{QML, QML2} followed by the softmax activation function. To achieve such a task, we first encode the input vector $\bm{x}_i$ into some quantum state $\rho_{\bm{x}_i}=\mathcal{C}(\bm{x}_i)\ket{0}^{\otimes n}\coloneqq\ket{\psi(\bm{x}_i)}$ via amplitude encoding to maximally utilize the degrees of freedom in the Hilbert space and thereby saving simulation memory, which only requires $n=\log_2(N)$ qubits for $N$ pixels. The state is evolved through a non-trivial unitary $U(\bm{\theta})$ with parameters $\bm{\theta}\in \left[-\pi,\pi\right]^D$, which is decomposed into a sequence of $D$ unitaries with parameter $\theta_{lm}$ at layer $l$ on qubit $m$
\begin{equation}
    U(\bm{\theta})=\prod\limits_{l}W_{l}\prod\limits_{m}e^{-i\theta_{lm}H_{lm}},
\end{equation}
where $\theta_{lm}$ and $H_{lm}=\hat{\bm{n}}_{lm}\cdot\bm{\sigma}$, $\bm{\sigma}=(X,Y,Z)^{T}$ are the training parameters for the single-qubit rotation gates $R_{lm}\coloneqq e^{-i\theta_{lm}H_{lm}}$, and $W_l$ is the entangling layer without parameters. For every $R_{lm}$, there are three degrees of freedom, which can be encoded into the Euler angles ($\alpha, \beta, \gamma$), 
\begin{equation}\label{eq: euler angle decomposition}
    R_{lm}=R_{Z}(\alpha_{lm})R_{Y}(\beta_{lm})R_{Z}(\gamma_{lm}).
\end{equation}
The circuit design is illustrated in \cref{fig: VQC}.

At the end of the circuit, to reduce the impact of ordinary barren plateaus \cite{local_cost_function_1}, the system is measured with respect to the single-qubit Pauli-$Z$ observables. We therefore define the loss function under the cross-entropy formalism such that
\begin{equation}
    \mathcal{L}_{\bm{\theta}}(\hat{\bm{y}}(\bm{x}_i), \bm{y}_i)=\sum\limits_{j=1}^{n}{\bm{y}_{ij}\log\left(\hat{\bm{y}}(\bm{x}_i)_j\right)},
\end{equation}
where $\hat{\bm{y}}(\bm{x}_i)$ are the probabilities of the image with label from 0 to 9 predicted by the model in the form of an array with each element $\hat{\bm{y}}(\bm{x}_i)_j=\text{softmax}\left(\text{Tr}(Z_j U(\bm{\theta})\rho_{\bm{x}_i}U(\bm{\theta})^{\dagger})\right)$. We also apply the activation \texttt{softmax} function to smooth the outputs,
\begin{equation}\label{eq: softmax definition}
    \text{softmax}\left(\text{Tr}\left(Z_j U_{\bm{\theta}}\rho_{\bm{x}_i}U_{\bm{\theta}}^{\dagger}\right)\right)=\frac{\exp(\text{Tr}\left(Z_j U_{\bm{\theta}}\rho_{\bm{x}_i}U_{\bm{\theta}}^{\dagger}\right))}{\sum\limits_{j'}\exp\left(\text{Tr}\left(Z_{j'} U_{\bm{\theta}}\rho_{\bm{x}_i}U_{\bm{\theta}}^{\dagger}\right)\right)}.
\end{equation}

\section{Numerical Results}\label{sec: Numerical results}
\subsection{Depolarizing channel on single-qubit gates}\label{subsec: depol channel}
Assuming errors in a general parameterized single-qubit gate persist, and that two-qubit gates are error corrected with a negligible logical error rate, we first introduce a depolarizing noise channel after every single-qubit unitary channel $\Lambda_i=R_{\hat{\bm{n}}_{lm}}(\theta_{lm})\bigcdot R^{\dagger}_{\hat{\bm{n}}_{lm}}(\theta_{lm})$ on qubit $i$ with strength $p$ to examine the feasibility of training those models with partial-QEC.  The combined channel is expressed as $\Tilde{\Lambda_i}\coloneqq \mathcal{E}_i\circ \Lambda_i$,
where
\begin{equation}\label{eq: depol channel kraus rep}
\begin{aligned}
    \mathcal{E}_i(\rho)&=(1-p)\rho+\frac{p}{3}\left(X_i\rho X_i+Y_i\rho Y_i+Z_i\rho Z_i\right),
    \\
    \Lambda_i(\rho)&=U_i\rho U_i^{\dagger}
\end{aligned}
\end{equation}

\begin{figure*}[t!]
    \centering
    \subfloat{
    \includegraphics[width=.484\linewidth]{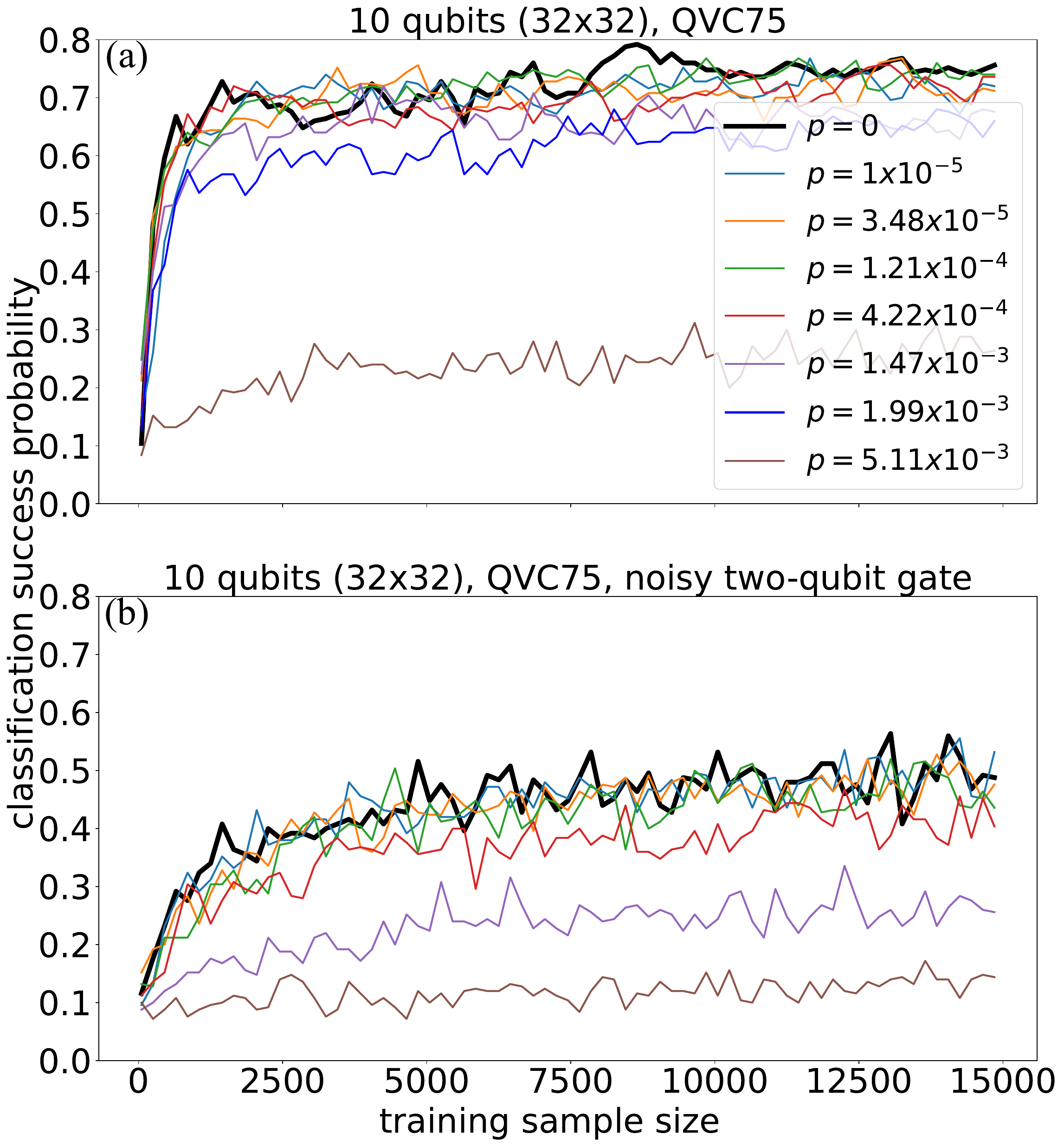}}
    \subfloat{\includegraphics[width=.516\linewidth]{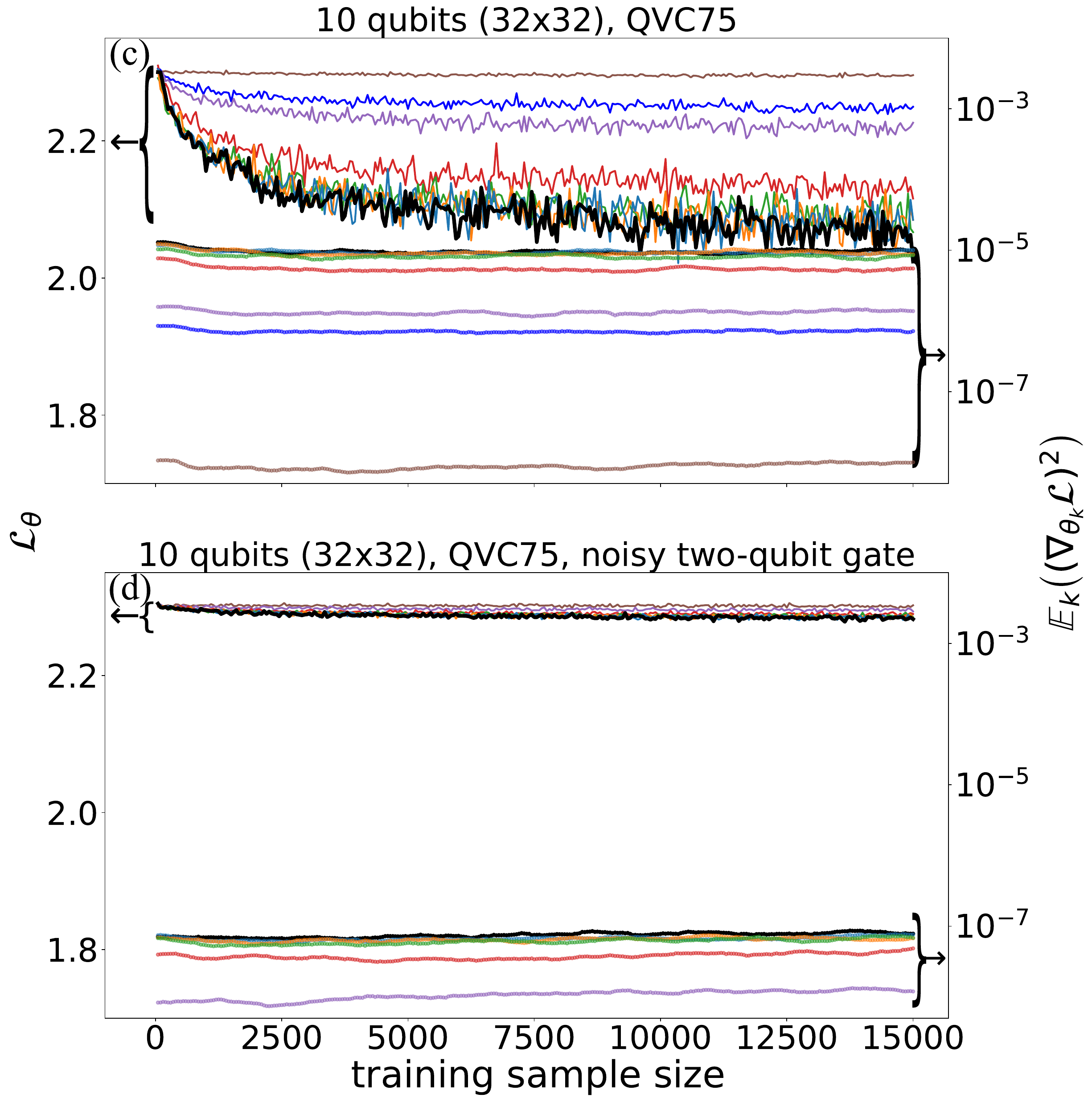}}
    \caption{The performance of the QVC in MNIST number classification problems with (a) 75 quantum layers but without a classical layer, and (b) 75 layers with a fully-connected classical layer and noisy two-qubit gate. The classification success rates inferred from the test datasets are plotted against the total number of images trained. The noise-free simulation is highlighted with black, thickened lines to contrast with noisy ones. Since we employ amplitude encoding, the number of pixels per image is exactly $2^n$, which corresponds to the dimension of $\sqrt{2^{n}}\cross \sqrt{2^{n}}$, where $n$ is the number of qubits for the variational circuit. The legend denotes the strengths of the depolarizing channel from 0 to $5.11\cross 10^{-3}$. (c) The corresponding cost function values (left axis) and their average gradients squared (right axis), $\mathbb{E}_k\left((\nabla_{\theta_k}\mathcal{L})^2\right)$, of the trainable parameters (classical and quantum), where the arrows indicate the axis correspondence of the plots. A clear trend of flattening loss landscapes and vanishing gradients can be observed as the depolarizing strength increases. When the gradients drop to a scale around $10^{-8}$, i.e. $p_{\text{depol}}=5.11\times10^{-3}$, due to shot noise, the model becomes distinctively not trainable. (d) For the noisy two-qubit gate plot where the classical layer is included, the averaged cost function gradients with respect to all parameters, including both the quantum and classical layers, actually increase after being trained with more images. In contrast, the classical layer gradients by themselves are behaving as expected, as shown in the zoomed figure.}
    \label{fig: depol plots long layers}
\end{figure*}
are the noise channel and unitary channel in the Kraus representation, respectively. Since the noise channel acts on the logical space, $p$ is interpreted as the combined logical $X$, $Y$, and $Z$-flip error rate. In practice, the channel would map the density operator after every decomposed gate of $R_{\hat{\bm{n}}_{lm}}$. However, since any unitary commutes with the depolarizing channel, it is convenient to treat it as in \Cref{eq: depol channel kraus rep}.

To evaluate the trainability of the variational circuit with only single-qubit noise, we used a learning rate of 0.005 and a total of 15000 images split into 50 images per batch in our training set, and 250 images in our test set, which are drawn and ordered randomly. The expectation values are evaluated on the Pauli-$Z$ operators, and every circuit is executed with 10000 shots to simulate real hardware with non-zero shot noise and avoid potential saddle points \cite{shot_noise_avoids_saddlepoints}. In some cases, we design and implement QVCC (QVC--classical) as illustrated in \Cref{fig: QVCC} (QVCC layout), where an additional fully connected classical layer $y_j=\sum\limits_{i}{w_{ij}}\expval{Z_i}+c_{ij}$
($i$ ranges from 1 to $n$ for $n$ qubits, $j$ ranges from 0 to 9 inclusive, and hence 10 outputs) is added after the quantum layers and before the softmax activation function to reshape the output to 10 neurons, matching the 10 possible classes, in case $n\ne10$. This extra classical layer can also serve to rescale the suppressed gradients with the purely quantum neural network. However, we expect the barren plateaus to remain, as the shot noise is also rescaled; therefore, the resolution on the Pauli-$Z$ operators is not improved.

\begin{figure}[t!]
    \centering
    \includegraphics[width=1\linewidth]{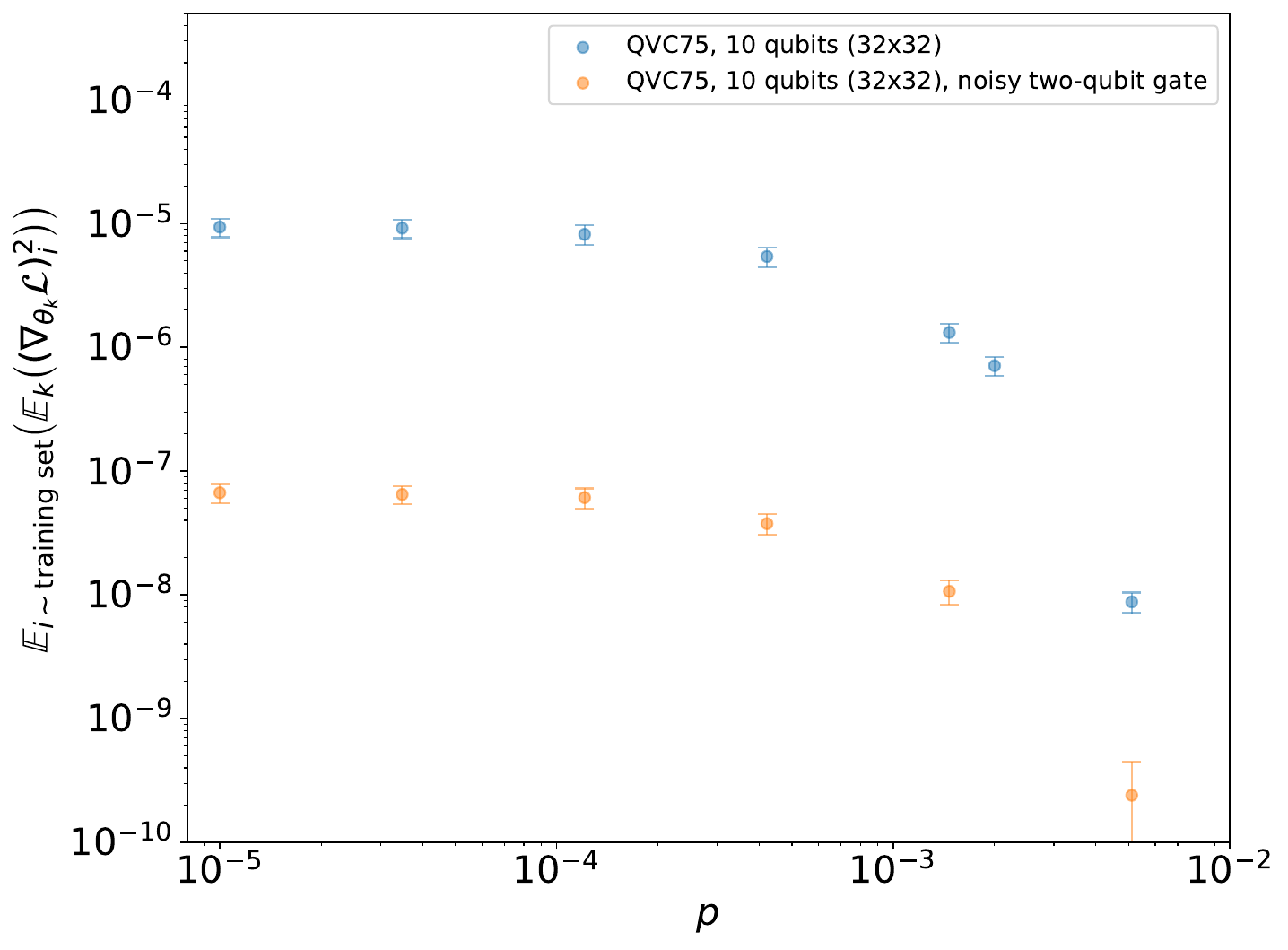}
    \caption{The gradients squared averaged over all trainable parameters and all iterations in the training versus the depolarizing strength $p$ in logarithmic scale. Error bar takes the standard deviation of the mean among all iterations.}
    \label{fig: gradients_vs_p_long_layers}
\end{figure}

As shown in \Cref{fig: depol plots long layers}a, the classification success rate attains a lower saturation more quickly when the depolarization strength increases. There is also a clear trend that both the gradients and the cost function landscape quickly flatten as the depolarizing noise becomes stronger, as is evident in \Cref{fig: depol plots long layers}c and \Cref{fig: gradients_vs_p_long_layers}. These results agree with the prediction of \cite{noise_induced_barren_plateaus}, reinforcing the argument that depolarizing noise induces barren plateaus, which have vanishing gradients that scale exponentially with their strength. In more supplementary plots \Cref{fig: classification_success_rate_short_layers,fig: loss_and_gradients_short_layers,fig: gradients_vs_p_short_layers} in \Cref{appendix: supplementary plots for depol channel}, we demonstrate that the above observations hold for QVC and QVCC of 50 and 100 layers, which exhibit higher (lower) gradients due to shorter (longer) circuit depths, as well as for different number of qubits, which exhibit higher (lower) gradients when there are more (less) qubits. 

Most importantly, as shown by the blue curves in \Cref{fig: depol plots long layers}a and \Cref{fig: depol plots long layers}c, despite shallower gradients and a smaller asymptotic success rate, the model is still trainable and above the shot noise at least for depolarizing strength of $p=1.99\cross 10^{-3}$ with only a slight reduction in the classification success rate. This corresponds to a logical parametrized single-qubit gate $U_{\hat{\bm{n}}_{lm}}(\theta_{lm})$ with net error rate of $1.33\times 10^{-3}$, a value can be achieved with underlying raw logical $T$ gate error rate no more than $\epsilon_T=10^{-4}$ by decomposing $U_{\hat{\bm{n}}_{lm}}(\theta_{lm})$ into Clifford + $T$ (See derivation in \Cref{appendix: Depolarising noise strength and actual gate error analysis} on conversion between depolarizing strengths to gate error rate through the randomized benchmarking protocol). Under the same logic, but applying \Cref{eq: randomized benchmarking fidelity,eq: gate error to fidelity} for two-qubit, an error-corrected logical two-qubit Clifford gate with an error rate of $3.20\times10^{-10}$, a value comparable to that estimated in \Cref{tab: full VQC costs}, only corresponds to a depolarizing strength of $p=1\times10^{-10}$. Given such a huge difference, it reinforces our approximation that the noise from the error-corrected two-qubit gates has a negligible contribution to the trainability of the model that involves noisy $T$ gates. Moreover, we note that even though a direct simulation of model performance with distilled gates $T$ was not presented in \Cref{fig: depol plots long layers}, the corresponding depolarizing strength would have been $p=7.38\times10^{-8}$ if a full-QEC, distilled circuit is implemented with logical $T$ gate error rate $\epsilon_T=2.47\times10^{-9}$ as estimated in \Cref{tab: full VQC costs}. This implies that it would fall into the range somewhere between $p=0$ and $p=1\times10^{-5}$ and much closer to $p=0$, the two specific legends in the plots that do not have a visible difference in performance. Therefore, the performance for the fully distilled gates would be very similar to the curves with $p=0$ in those plots. We further analyze the trainability of the model by investigating the upper bound of the variation in the loss function caused by the statistical fluctuations of the expectation values (See detailed analysis in \Cref{appendix: variation in loss caused by shot noise}). We confirm that for the case without a classical layer, the average gradient above $6\times10^{-3}$ is guaranteed to be trainable in our setting, which is in agreement with our observation in \Cref{fig: depol plots long layers,fig: loss_and_gradients_short_layers}.

Furthermore, as shown in \Cref{fig: depol plots long layers}b and \Cref{fig: depol plots long layers}d, we test the QVC model performance when the two-qubit gate noise is also turned on, which is aimed at simulating when the circuit is purely executed on physical qubits without error-correction, thereby demonstrating the necessity of partial QEC in balancing the model's trainability and overhead compared to no QEC. Similarly to the single-qubit gate noise channel, the two-qubit gate noise is added after the noise-free unitary: $\Tilde{\Lambda_i}\coloneqq \mathcal{E}_{(q_1,q_2)}\circ \Lambda_{(q_a,q_b)}$, where $q_1$, $q_2$ are the two qubits the gate is acting on. The overall noise channel $\mathcal{E}_{(q_1,q_2)}$ contains two subchannels: depolarizing channel $\mathcal{E}_{\text{depol}_i}$ on qubit $i$ with the same structure as in \Cref{eq: depol channel kraus rep}, and crosstalk nearest-neighbour $ZZ$ couplings $\mathcal{E}_{\text{crosstalk}_{(q_1,q_2)}}$ on qubits $q_1$ and $q_2$, where
\begin{equation}
    \mathcal{E}_{\text{crosstalk}_{(q_1,q_2)}}[\bigcdot] = U_{Z_{q_1}Z_{q_2}}\bigcdot U^{\dagger}_{Z_{q_1}Z_{q_2}}.
\end{equation}
The unitary $U_{Z_{q_1}Z_{q_2}} = \exp\left(-i\alpha Z_{q_1} Z_{q_2}\right)$ is simply a two-qubit unitary with generating Hamiltonian $ZZ$ and angle of rotation $2\alpha$. Since multiple channels are involved, we describe each channel $\mathcal{E}_i \coloneqq e^{\mathcal{L}_i}$ with Lindbladian generator $\mathcal{L}_i$ in the superoperator representation, where
\begin{equation}
    \mathcal{E}_i=\sum\limits_{j}K_{ij}^{T}\otimes K_{ij},
\end{equation}
$K_{ij}$ are the Kraus operators of channel $i$. Therefore, the net Lindbladian
\begin{equation}
\begin{aligned}
\mathcal{L}&=\mathcal{L}_{\text{depol}_{q_1}} + \mathcal{L}_{\text{depol}_{q_2}} \\
    &+ \mathcal{L}_{\text{crosstalk}_{(q_1-1,q_1)}} + \mathcal{L}_{\text{crosstalk}_{(q_2,q_2+1)}}
\end{aligned}
\end{equation}
is the sum of the Lindbladian generator of all noise channels for the master equation,
\begin{equation}
    \frac{d\rho}{dt}=\mathcal{L}\left[\rho\right],
\end{equation}
and have the solution $\vert\rho(t)\rrangle= e^{\mathcal{L}t}\vert\rho(0)\rrangle$, where $\vert\rho\rrangle$ denotes the density operator in vectorized form.
For a more realistic simulation, the depolarizing and crosstalk noise strengths are characterized by the IBM superconducting quantum computer $ibm\_torino$. From the official average two-qubit gate error rate of $3.8\cross 10^{-3}$ characterized based on randomized benchmarking, we derive $p_{\text{depol}}=0.0019$. A single-qubit example of this derivation is discussed in \Cref{appendix: Depolarising noise strength and actual gate error analysis}. The strength of the crosstalk $ZZ$ couplings, $\alpha=1.16\cross 10^{-3}$, is taken from the noise parameters obtained from previous work \cite{haiyue_crosstalk}.

When two-qubit gate noises are added back, the classification success rate is much worse. In all situations, the classification success rate never crosses 0.5 and the average squared gradient  $\mathbb{E}_k\left((\nabla_{\theta_k}\mathcal{L})^2\right)$ is below $10^{-7}$. At the single-qubit depolarizing channel strength of $p=1.47\cross 10^{-3}$, classification success rate barely increases and the model is almost non-trainable, whereas the model behaves properly for the case without the two-qubit gate noises. The comparison between noisy and noise-free two-qubit gates demonstrates the utility of partial QEC, which significantly improves the trainability of the machine learning model while still achieving a drastic overhead reduction by forgoing distillations. 
\begin{figure*}[t!]
    \centering
    \subfloat{\includegraphics[width=.485\linewidth]{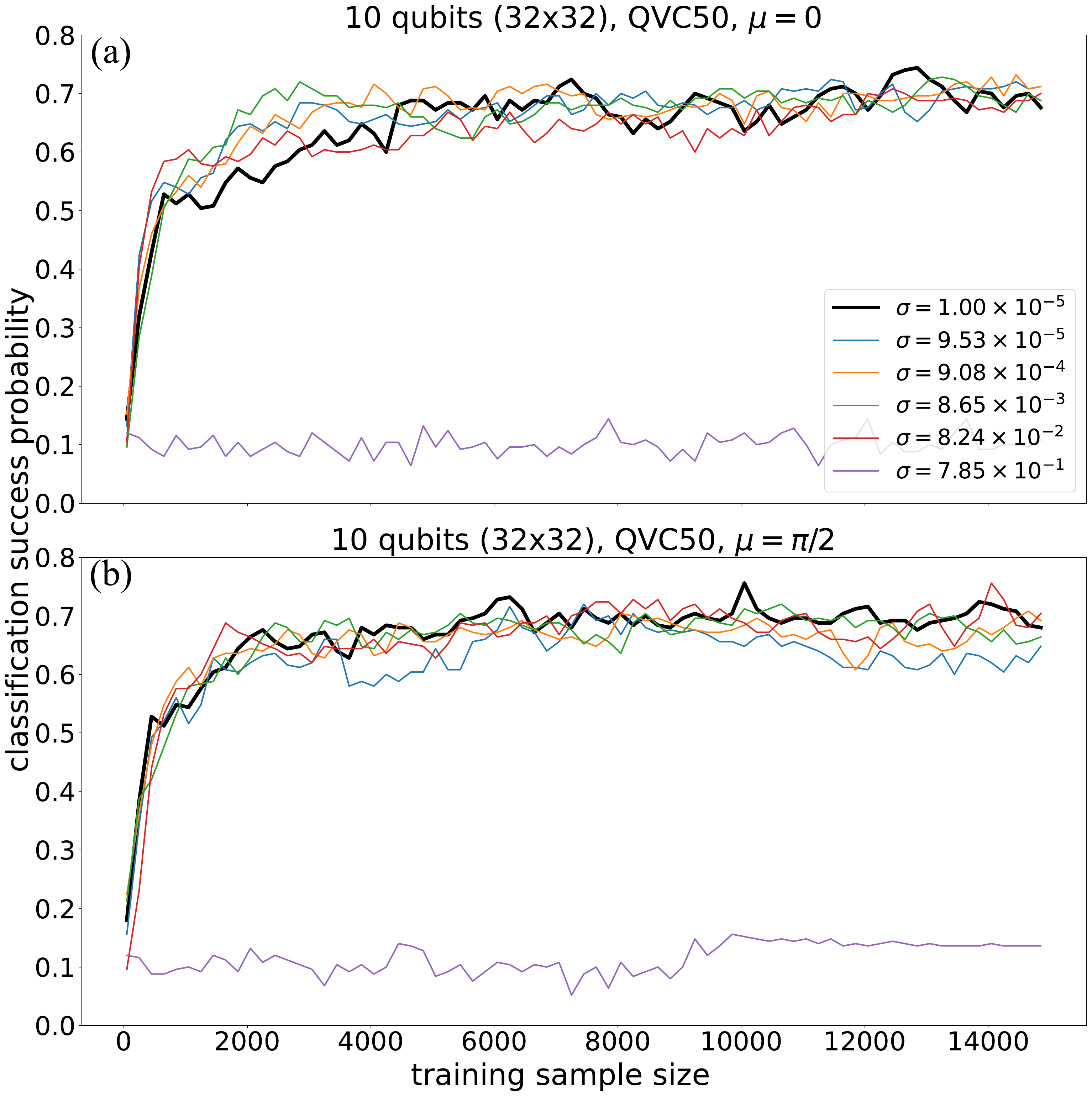}}
    \subfloat{\includegraphics[width=.514\linewidth]{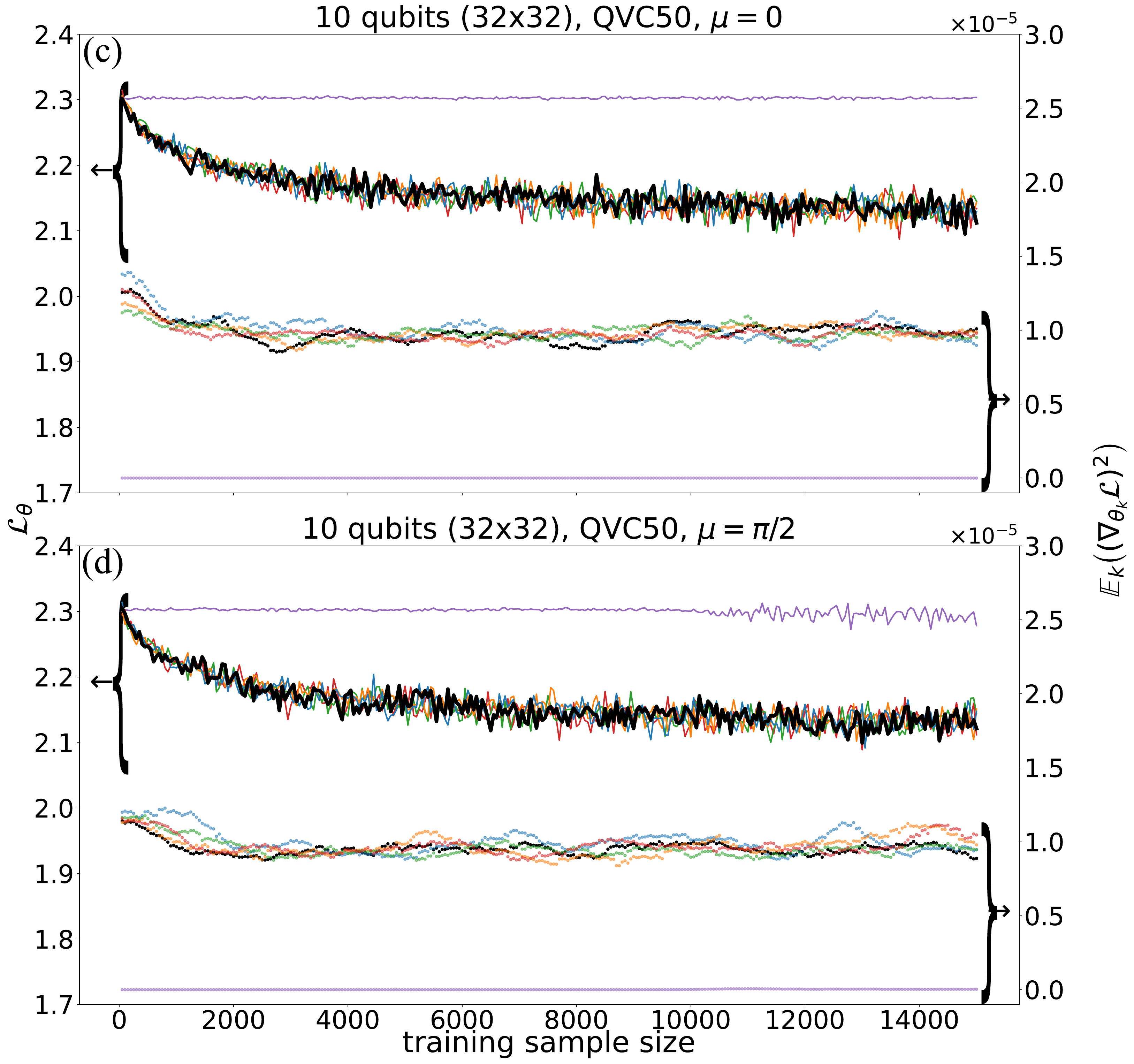}}
    \caption{(a) and (b) are the classification success rates, (c) and (d) are the cost function and its average squared gradients $\mathbb{E}_k\left((\nabla_{\theta_k}\mathcal{L})^2\right)$ plotted together (arrows indicate axis belongings) for when the phase-damping channel is introduced after the single-qubit gates. Simulations with the smallest phase damping standard deviation $\sigma$ are highlighted with black, thickened lines. The plotting structures and notations are exactly the same as \Cref{fig: depol plots long layers}, except for the legends, which correspond to the standard deviations of the phase-damping channel spread. The hyperparameter $\mu$ indicates the mean of the density function \Cref{eq: pdf function} that controls the mean angle of over-rotation. Notably, the behavior of the training curves of the model is not affected when the mean angle is non-zero.}
    \label{fig: gaussian phase damping noise plots}
\end{figure*}
\subsection{Phase damping channel with non-zero means on single-qubit gates}
In practice, depolarization is not the only noise channel that affects a quantum state, even in the logical space: Multiple errors corresponding to the same syndrome may occur with different probabilities, causing the noise to be more directional in the logical Bloch sphere and behave more closely to a unitary compared to the depolarizing channel, at least before the final readout. In that case, we expect the QNN model to demonstrate improved performance in trainability compared to the full depolarizing model, which suffers the most from noise-induced barren plateaus \cite{noise_induced_barren_plateaus} that lead to a complete classical random mixture with maximal entropy. Consequently, we introduce the phase-damping channel, which simulates the over- or under-rotation of an ideal single-qubit rotation, thus constituting the ideal gate $\Lambda_i$, along with the noise channel:
\begin{widetext}
    \begin{equation}\label{eq: phase damping channel}
    \mathcal{E}_i(\rho)=\int\limits_{-\infty}^{\infty}d\theta P(\theta)R_{Z}(\theta)_i\rho R_{Z}(\theta)^{\dagger}_i=\left(\int\limits_{-\delta}^{\delta}+\int\limits_{-\infty}^{-\delta}+\int\limits_{\delta}^{\infty}\right)d\theta P(\theta)R_{Z}(\theta)_i\rho R_{Z}(\theta)^{\dagger}_i=I_{\delta}+I_{-\infty}+I_{\infty}
\end{equation}
\end{widetext}
where $\theta$ is the angle of over- or under-rotation, and
\begin{equation}\label{eq: pdf function}
    P_{\mu=0}(\theta)=\frac{1}{\sqrt{2\pi\sigma^2}}e^{-\theta^2/2\sigma^2}
\end{equation}
is the Gaussian distribution with zero mean and standard deviation $\sigma$. $\delta$ is some arbitrarily small value. In the limits where $\delta/\sigma\gg1$, this is approximated and simplified to 
\begin{equation}\label{eq: gaussian channel at low sigma limit}
    \mathcal{E}_i(\rho)=\left(1-\Phi(\delta)\right)\rho+\frac{\Phi(\delta)}{2}\{R_{Z}(\delta)_i\rho, R_{Z}(\delta)^{\dagger}_i\}
\end{equation}
where $\Phi(\delta)=e^{-\left(\frac{\delta}{\sigma}\right)^2}\left((\delta/\sigma)^{-1}+O\left(\left(\delta/\sigma\right)^{-3}\right)\right)$ is the Gaussian CDF series expansion from $\delta\rightarrow\infty$. Hence, this is almost a unitary (identity) evolution. Even for relatively larger $\sigma$, unlike Pauli channels, most of the underlying evolutions $R_{Z}(\theta)_i\rho R_{Z}(\theta)^{\dagger}_i$ are still very close to the original state $\rho$, making the image of the channel almost a pure state. Unlike \Cref{subsec: depol channel}, where all depolarizations are commuted to the end, the phase damping channel is applied after the decomposed gate as in \Cref{eq: euler angle decomposition} for a more realistic simulation.

We also consider the possibility that the Gaussian distribution $P_{\mu}(\theta)$ has a non-zero mean. In application, this corresponds to the standard phase-damping channel plus a non-zero idling background Hamiltonian $H=\mu\frac{Z}{2}$, and the evolution is described as
\begin{equation}
    \frac{d\rho}{dt}=-i\left[H,\rho\right]+\mathcal{L}_{\text{phase-damping}}\left[\rho\right].
\end{equation}
In such a situation, it is clear that $\mathcal{E}_i(\rho)_{\mu\ne0}=\mathcal{E}_i(R_{Z}(\mu)\rho R_{Z}(\mu)^{\dagger})_{\mu=0}$. However, we expect only a minimal or even zero impact of the non-zero mean on the performance of a QVC model compared to the zero mean case, due to the fact that it is completely possible for the model to `learn' this mean and compensate for it by applying the opposite rotation:
\begin{equation}\label{eq: non zero mean overrotation}
\begin{aligned}
    &(\mathcal{E}\circ \Lambda)_{\mu\ne0}(\rho)\\&=\mathcal{E}_{\mu=0}(R_{Z}(\mu)\Lambda_{\mu\ne0}(\rho) R_{Z}(\mu)^{\dagger})\\
    &=\mathcal{E}_{\mu=0}(R_{Z}(\mu)R_{Z}(-\mu)\Lambda_{\mu=0}(\rho) R_{Z}(\mu)^{\dagger}R_{Z}(-\mu)^{\dagger})\\
    &=\mathcal{E}_{\mu=0}(\Lambda_{\mu=0}(\rho) )\\
    &=(\mathcal{E}_{\mu=0}\circ \Lambda_{\mu=0})(\rho),
\end{aligned}
\end{equation}
where we define $\Lambda_{i,\mu\ne0}=R_{\hat{\bm{n}}}(-\mu)\circ\Lambda_{i,\mu=0}$ that can be learned by the model during its training.

As evident in \Cref{fig: gaussian phase damping noise plots}, the QVC model has demonstrated strong robustness against the phase damping channel in its training process, where its classification success rate, gradients, and its trainability is almost unperturbed when the standard deviation of \Cref{eq: pdf function} is less than $8.24\times10^{-2}$. This supports our prediction in \Cref{eq: gaussian channel at low sigma limit} of the Gaussian phase damping channel at the small $\sigma$ limit (assuming that $\delta$ is fixed) that it is almost an identity mapping. However, once the standard deviation crosses a threshold, the model quickly becomes non-trainable. Meanwhile, the efficiency of the training is also unaffected when a significant non-zero mean of the over-rotation, namely $\mu=\frac{\pi}{2}$ is implemented, as predicted according to \Cref{eq: non zero mean overrotation}. This implies that in a practical situation with a biased background Hamiltonian that induces idle over-rotation, a variational circuit model can adapt to the noise through training, or equivalently, the analogy of noise-induced barren plateaus for unitary rotations does not hold.
\begin{figure}[t!]
    \centering
\includegraphics[width=\linewidth]{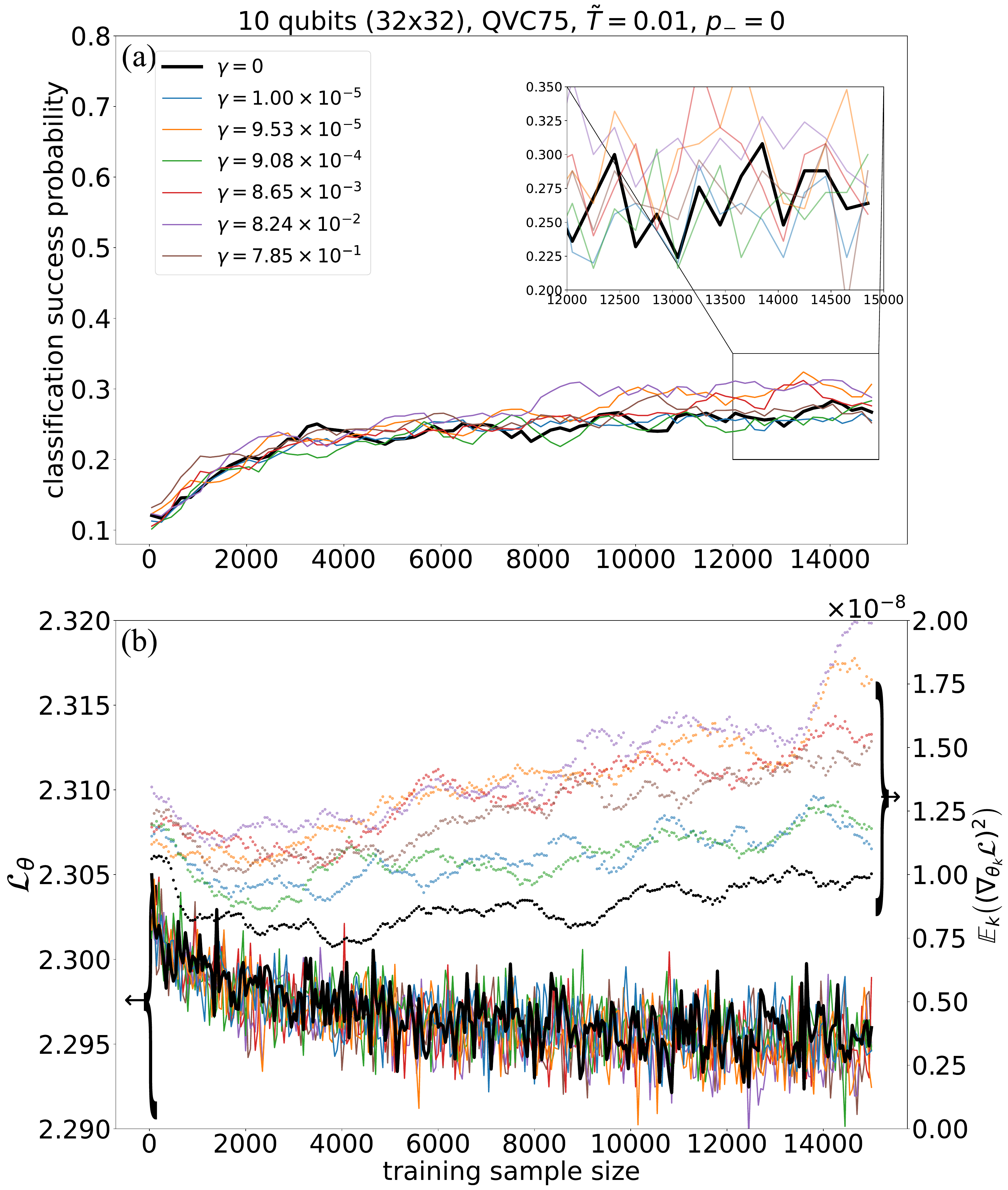}
\caption{(a) The classification success rate when the depolarizing channel and thermal damping channel are both involved. 10 qubits with 75 QVC layers are used, and the generalised temperature $\Tilde{T}$ is set to be 0.01. Each coloured plot represents one decay rate $\gamma$ from 0 to $7.85\times 10^{-1}$ in the logarithmic scale, where $\gamma=0$ indicates no thermal damping and is highlighted with black thickened lines to contrast with others. Note that to better visualise any tiny differences in the general trend, we average over the nearest 5 test batches to smooth the curve. The zoomed-in figure depicts the actual sampled success rate without smoothing. Clearly, after adding the thermal damping channel at low temperatures, the classification success rate has been improved, and the (b) average squared gradients $\mathbb{E}_k\left((\nabla_{\theta_k}\mathcal{L})^2\right)$ of the loss function have been improved, driving the model away from noise-induced barren plateaus. Arrows indicate the axis to which the data belong.}
\label{fig: thermal damping noise plots}
\end{figure}
\subsection{Thermal damping channel at low temperature}
Although we have shown that it is sufficient to train a QVC model with partial QEC even under considerable depolarizing noise, further mitigation would still be beneficial. In this subsection, we investigate the possibility of adapting thermal damping noise as a countermeasure against the depolarizing noise, and further reinforce the feasibility of applying partial-QEC in the training of a QVC model. The intuition comes from the fact that, unlike a depolarizing channel that mixes pure quantum states, the thermal damping channel purifies them at low temperatures. In fact, proof-of-concept works have already been demonstrated in literature \cite{thermal_helps_learning,thermal_helps_learning2}. 

We model the noise channel $\mathcal{E}_i$ acting after the noise-free gate $\Lambda_i$ as usual, and its evolution on a single-qubit state is described as
\begin{equation}
    \frac{d\rho}{dt}=\mathcal{L}_{\text{depol}}\left[\rho\right]+\sum\limits_{i=\pm}{\Gamma_i\left(\sigma^{\dagger}_i\rho\sigma_i-\frac{1}{2}\{\sigma_i\sigma_i^{\dagger},\rho\}\right)},
\end{equation}
where $\sigma_{-}=\ket{0}\bra{1}=\sigma_{+}^{\dagger}$ is the lowering operator, $\Gamma_{\pm}=\frac{\Gamma_0 e^{\pm\frac{\hbar\omega}{2k_BT}}}{\sinh\left(\frac{\hbar\omega}{2k_BT}\right)}$ are the decay and excitation rates with spontaneous emission rate of $\Gamma_0$. In Kraus-operator form, we have $\mathcal{E}_{\text{thermal}}(\rho)=\sum_{i}E_i\rho E_i^{\dagger}$, $\sum_i E_i^{\dagger}E_i=\mathds{1}$, with
\begin{equation}
\begin{aligned}
    E_0&=\sqrt{p_-}\left(\ket{0}\bra{0}+\sqrt{1-\gamma}\ket{1}\bra{1}\right)\\
    E_1&=\sqrt{p_-\gamma}\sigma_{-}\\
    E_2&=\sqrt{1-p_-}\left(\sqrt{1-\gamma}\ket{0}\bra{0}+\ket{1}\bra{1}\right)\\
    E_3&=\sqrt{(1-p_-)\gamma}\sigma_+,
\end{aligned}
\end{equation}
where $p_-=\frac{e^{\frac{\hbar\omega}{2k_BT}}}{2\cosh\left(\frac{\hbar\omega}{2k_BT}\right)}$ is the probability of spontaneous decay, $\gamma=1-e^{-\Gamma t}$ is the decay rate for the elapsed time $t$ and $\Gamma=\Gamma_+ +\Gamma_-$.

In our simulation, we set the generalised temperature $\Tilde{T}\coloneqq \frac{2k_BT}{\hbar\omega}=0.01$, which makes thermal relaxation act as an amplitude damping channel from $\ket{1}$ to $\ket{0}$ state. Therefore, we can approximate $\Gamma\approx \Gamma_0$, hence $\gamma\approx1-e^{-\Gamma_0 t}$. We also set the strength of the depolarizing channel to be $p=5.11\times10^{-3}$, as it corresponds to the poorest performance in the previous simulation, as in \Cref{fig: depol plots long layers}. In order to observe the impact of the thermal damping channel purely and directly on the quantum observable, the classical layer is not activated. To evaluate the performance at different damping strengths, we scan $\Gamma_0t$ from $1.0\times10^{-6}$ to $1.0\times10^{-1}$ on the logarithmic scale, which corresponds to the decay rate from $\gamma=1.0\times10^{-7}$ to $\gamma=9.5\times10^{-2}$. As shown in \Cref{fig: thermal damping noise plots}b, the higher the decay rate, the more significant the increase in gradients compared to $\gamma=0$ without thermal damping. In particular, when thermal damping is activated, we note that the gradients no longer decay as more samples are trained, while the loss function approaches its minimum. However, when only the depolarizing channel is present, overall the gradients are suppressed. This demonstrates the effectiveness of the thermal damping channel in playing a positive role against noise-induced barren plateaus by improving the average size of the gradients. We also remark that the classification success rate is indeed higher when the thermal damping decay rate is larger, but is only distinguishable when the training sample size is large and the classification success rate saturates. Nevertheless, it still outperforms the pure depolarization noise model when the damping decay rate is large enough ($\gamma=8.65\times10^{-3}$), as the model was already mostly trainable even when only depolarization is present.

\section{Discussion}\label{sec: Discussion}
To address the issue of noisy quantum gates for the application of supervised quantum machine learning that suffers from noise-induced barren plateaus, we proposed a partial quantum error correction protocol. By forgoing the magic state distillation, we balanced the conflicting needs of minimal mitigation overhead and effectiveness against depolarizing noise. Through a literature review and careful analysis of the surface code cost of the $T$ factory and injection, we remark that the spacetime complexity to conduct a single instance of distillation is at least 11 times or 14600 qubit-rounds more expensive compared to using the raw magic state directly. We also derived an analytical expression for its spacetime cost for an arbitrary number of distillation layers. The simulation tells that, to achieve a reasonable logical error rate, the cost of the whole variational circuit differs by two orders of magnitude, depending on whether magic state distillation is implemented.

Given this significant overhead reduction, which makes full fault-tolerant QML implementation infeasible in the foreseeable future, we simulated the quantum variational classifier (QVC) and QVCC (QVC with additional classical layer) models' performance, particularly their trainability, under single-qubit gate noise only. In the experiments for classifying the handwritten numbers of the MNIST datasets, we presented clear evidence of the model's trainability with a depolarizing noise strength of $p=1.47 \times 10^{-3}$, which corresponds to a gate error rate of $1.96 \times 10^{-3}$, a value higher than that of state-of-the-art quantum computers. However, this does not deny the overall trend predicted by the noise-induced barren plateaus as the noise becomes stronger. Moreover, the poor performance when both single- and two-qubit noise was turned on further reinforces the necessity of partial quantum error correction instead of directly running on physical qubits. We further note the robustness of the variational circuit model under phase-damping noise, simulating a non-zero background Hamiltonian that can be adapted during the training process. In terms of the noise channels other than depolarizations, our simulation demonstrates that the QVCC model can completely adapt and is therefore unaffected by the mean angle of over-rotation in Gaussian phase-damping noise. When a thermal damping channel is added at low temperature, the trainability of the QVC model has been improved, further verifying the hypothesis that thermal damping noise can play a positive role in quantum machine learning \cite{thermal_helps_learning,thermal_helps_learning2}. Quantum processors are being developed by many hardware companies and research institutes around the world; however, in the next few years, the size of quantum processors is expected to be incompatible with the QEC resource requirements. Our work has proposed the idea of partial QEC for the implementation of QML models, which will be feasible on the near to medium-term quantum devices. These partially corrected QML models will be free from trainability issues and demonstrate resilience to hardware noise even for deep quantum circuits necessary for practical applications.  

\bigskip
\noindent
\textbf{Further notes}\par
\noindent
After submitting our work to arXiv, previous work on partial QEC in Refs. \cite{partial_QEC_direct_R1,partial_QEC_direct_R2} was brought to our attention, which demonstrated a way to implement logical $R_{Z}(\theta)$ via direct state injection rather than breaking down into Cliffords and $T$ gates. However, the error rate of their idea was only benchmarked at the gate level and did not demonstrate algorithmic-level improvement by performing simulations or by including the QVC implementations. We not only propose the idea of partial QEC in the context of quantum algorithms, our work is also the first to demonstrate its implementation and working for large-depth QVC model scenarios.
\begin{acknowledgements}
    The authors greatly acknowledge the computational resources provided by the National Computing Infrastructure (NCI) through the National Computational Merit Allocation Scheme (NCMAS), and the University of Melbourne's Research Computing Services. H.K. was supported by the Australian Government Research Training Program Scholarship.\\
\end{acknowledgements}

\noindent
\textbf{Data Availability Statement}\par
\noindent
The source code and raw data used in the current study will be openly available at \cite{partial-QEC-QVC}.\\

\noindent
\textbf{Author Contributions}\par
\noindent
MU conceived the original idea. MU and MS supervised the project. HK developed the QML simulation framework and carried out all experiments. EA, HK and YK computed estimates of QEC resources with input from MU. HK, MU, and MS analyzed the data. HK wrote the manuscript with the input from all authors.

\appendix
\section{Estimate of full variational circuit cost}\label{appendix: Cost of full variational circuit}
With the assistance of Microsoft Azure quantum resource estimator \cite{Azure_quantum_resource_estimator, Azure_quantum_resource_estimator_paper} based on defect-based logical qubits introduced in Ref. \cite{surface_codes}, we simulate the total number of physical qubits and the runtime needed to implement a full variational circuit with or without distillations. Although this only demonstrates the behavior for defect-type surface codes, the logic still applies to other surface codes, such as the patch-type ones.
Specifically, given a physical gate error rate of $10^{-3}$, which includes both the physical Clifford gates and the raw magic state, the costs to achieve a full-circuit error budget of $10^{-3}$ and $10^{-4}$ are summarized as in \Cref{tab: full VQC costs}. We note that the full-circuit error budget $\epsilon$ is equal to
\begin{equation}
    \epsilon=\epsilon_{\text{log}}+\epsilon_{\text{dis}}+\epsilon_{\text{syn}},
\end{equation}
for contributions from logical, distillation, and synthesizing errors (the error of synthesizing rotation gates with arbitrary angles), which are equally distributed in our simulation. As a result, the required average logical error rate per qubit per stabilizer cycle $\epsilon_L$ or distilled $T$ gate error rate $\epsilon_T$ must be such that it does not exceed its contribution to the error budget,
\begin{equation}\label{eq: error budget to logical error rate}
    1-(1-\epsilon_{L})^{N_L}\le\epsilon_{\text{log}}\text{ or } 1-(1-\epsilon_{T})^{N_T}\le\epsilon_{\text{dis}},
\end{equation}
where $N_T$ is the corresponding number of logical $T$ gates, $N_L$ total number of patch-cycles. In the limit where $\epsilon_L\ll1$, \Cref{eq: error budget to logical error rate} approximates to 
\begin{equation}
    N_L\epsilon_L\le \epsilon_{\text{log}}\text{ or }  N_T\epsilon_T\le \epsilon_{\text{dis}}.
\end{equation}
As a result, the maximum allowed logical qubit error rate per logical qubit per cycle is
\begin{equation}
    \max_d\left(\epsilon_L=0.03\left(\frac{p}{p^{*}}\right)^{\frac{d+1}{2}}\right)
\end{equation}
that also satisfies \Cref{eq: error budget to logical error rate}.  We omit the next lower integer on $\lfloor\frac{d+1}{2}\rfloor$ as we always assume $d$ is odd. $p$ is the maximum physical error rate per syndrome round, $p^{*}=0.01$ is the physical error rate threshold below which the error rate of the logical qubit is less than the error rate of the physical qubit, and 0.03 is the coefficient extracted numerically from simulations when fitting an exponential curve to model the relationship between the logical and physical error rate. We remark that, due to the discretized nature of the allowed logical error rate per qubit per cycle, the actual full-circuit failure probability is always less than the benchmarked error budget. 

From the simulation, the overhead of QEC for the variational circuit is dominated by the $T$ factory that produces magic states for distillation. Otherwise, we estimate the resources spent to be comparable to those spent on data qubits, which scales in $O\left(\left(2Q_{\text{alg}}+\sqrt{8Q_{\text{alg}}}+1\right)d^2\right)$. This is because:
\begin{itemize}
    \item $T$ factory has a code distance $d$ same as the data qubits.
    \item There are at most $Q_{\text{alg}}$ logical $T$ gates can be executed in parallel. Therefore, a maximum of $Q_{\text{alg}}$ magic states need to be prepared simultaneously.
    \item Since distillation is no longer present, the time needed to prepare a magic state is no more than the minimum number of rounds ($\sim d$) for syndrome extraction, and therefore it is always prepared faster than it can be injected. As a result, no redundant magic states are needed even if there are $Q_{\text{alg}}$ magic states injected into every possible layer.
\end{itemize}

According to the simulation, the time cost is largely a linear relationship with respect to the number of QVC layers. This should still be the case without distillation, yet with a significantly smaller time cost per logical patch-cycle.

\section{Spacetime cost of magic state distillation}\label{appendix: Spacetime cost of magic state disillation}
\begin{table*}[t!]
    \centering
    \resizebox{\textwidth}{!}{
    \begin{tabular}{c|c|c|c|c}
         \toprule & & \textbf{Space cost} ($d^2$) & \textbf{Time cost} ($d$) & $p_{\text{distilled}}$ \\\hline
         \multirow{2}{*}{\shortstack{Single magic\\ state \cite{magic_state_distillation_cost1}}}
         &single layer & $\frac{1.5(m_X+k)+4}{k}$  & $\frac{n-m_X}{(1-p)^n}$  & $p_1=C p^{\frac{d+1}{2}}$ \\\cline{2-5} 
         &$L$ layers & $\left(\frac{1.5(m_X+k)+4}{k}\right)^L$ & $(n-m_X)\sum\limits_{l=1}^{L}{\left((1-p_l)^{\left(\frac{n}{k}\right)^{L-l} n}\right)^{-1}}$  & $p_l=C^{\frac{(d+1)(l-1)l}{2}}p^{\left(\frac{d+1}{4}\right)^l}$  \\\hhline{=|====}
         \multirow{4}{*}{\shortstack{Single magic\\
         state \cite{magic_state_distillation_cost2}}}
         &(15-to-1)$_{(7,3,3)} (d=11)$ &  6.69  & 1.65 & $4.4\cross 10^{-8}$ ($p=10^{-4}$) \\\cline{2-5}
         &(15-to-1)$_{(9,3,3)} (d=13)$ & 4.51 & 2.78 & $1.5\cross10^{-9}(p=10^{-4})$\\\cline{2-5}
         & (15-to-1)$^{4}_{(9,3,3)}\cross$(20-to-4)$_{(15,7,9)} (d=19)$& 45.43 & 4.75 & $2.4\cross 10^{-15}(p=10^{-4})$\\\cline{2-5}
         &No distillation & 1 & 1 & $p$
         \\\bottomrule
    \end{tabular}
    }
    \caption{Brief summary on the cost of distillation protocols in Ref. \cite{magic_state_distillation_cost1, magic_state_distillation_cost2} (details discussed in \Cref{appendix: Spacetime cost of magic state disillation}). $m_X$ is the number of $X$ stabilizers, $n$ is the number of magic states consumed, and $k$ is the number of magic states distilled. $p_l$ is the probability of having incorrect parities at layer $l$, or the error rate of the distilled $T$ gate. $p_l$ is a function of $p$, $C$, and $d$, which are the initial logical $T$ gate error rate without any distillation, a constant that depends on the number of parity combinations that would lead to a logical error, and $d$ is the code distance, respectively. If without any distillation, the spacetime cost for injection simply scales in $O(1)$ patch-cycles.}
    \label{tab: distillation costs}
\end{table*}

Based on the cost of a single layer of distillation as described in Ref. \cite{magic_state_distillation_cost1}, we derive the general formula for the spatial and temporal cost of the magic state distillation after $L$ layers, and summarize these in \Cref{tab: distillation costs},

First, from Eq. 9 in Ref. \cite{magic_state_distillation_cost1}, the number of patches needed to be prepared is
\begin{equation}
    \frac{1.5(m_X+k)}{k}
\end{equation}
per every distilled state, which has $d^2$ data qubits per patch. $m_X$ is the number of $X$ stabilizers, $k$ is the number of states distilled by consuming $n$ logical $T$ states from previous iteration. Therefore, it can be easily shown that after $L$ layers, the spatial cost is simply
\begin{equation}
     C_{\text{spatial}}(L)=\left(\frac{1.5(m_X+k)}{k}\right)^L.
\end{equation}
On the other hand, the number of syndrome measurement rounds needed, proportional to $d$ is given by
\begin{equation}
    \frac{n-m_X}{(1-p)^n},
\end{equation}
where $p$ is the logical error rate of the $T$ gate injected from the previous iteration. As a result, the total number of cycles for $L$ layers of distillation is simply the sum of the cycles for each layer:
\begin{equation}
    C_{\text{temporal}}(L)=\sum\limits_{l=1}^{L}{C_{\text{temporal, layer } l}}.
\end{equation}
Here, the time to execute the logical state injection circuit is set to $n-m_X$ if no wrong-parity stabilizers are measured. However, the actual syndrome rounds needed are damped according to the probability of measuring the wrong parity, which must be discarded, and distillation is restarted. Therefore, the probability of successfully passing the $l^{\text{th}}$ layer of distillation is given by
\begin{equation}\label{eq: l layer success rate}
    P(l^{\text{th}} \text{ success}\vert l'<l^{\text{th}} \text{ succeeded})=(1-p_l)^{n\left(\frac{n}{k}\right)^{L-l}},
\end{equation}
where $p_l$ is the purified logical error rate up to the $l^{\text{th}}$ layer, and $(1-p_l)^n$ is the probability of successfully distilling a magic state, which consumes $n/k$ magic states from the $(l-1)^{\text{th}}$ layer. Therefore, there are a total of $\left(\frac{n}{k}\right)^{L-l}$ lots of states distilled in the $l^{\text{th}}$ layer, hence having the probability of success given by \Cref{eq: l layer success rate}. On top of this, we also have
\begin{equation}
    p_{l+1}=Cp_l^{\frac{d+1}{2}},
\end{equation}
where $C$ is some constant depending on the number of false parities. Therefore, by induction,
\begin{equation}
    p_l=C^{\frac{(d+1)(l-1)l}{2}}p_0^{\left(\frac{d+1}{4}\right)^l},
\end{equation}
where $p_0$ is the initial \textit{logical} error rate without any distillation, the power of $C$ is deduced from the arithmetic summation from 0 to $l^{\text{th}}$ layer. We note that $p_0$ is not the same as the \textit{physical} single-qubit $\ket{T}$ state error rate. Instead, it is
\begin{equation}
    p_0=\frac{2}{5}p_{\text{2-qubit}}+\frac{2}{3}p_{\text{1-qubit}}+2p_I+O(p_{\text{phy }\ket{T}}^2),
\end{equation}
where $p_{\text{1-qubit}}$, $p_{\text{2-qubit}}$, $p_I$, and $p_{\text{phy }\ket{T}}$ are the single-qubit physical gate, two-qubit physical gate, initialization (preparation), and physical $\ket{T}$ state error rate, respectively \cite{physical_T_and_logical_T_error_rate}. It is apparent that the contribution from the physical $\ket{T}$ state is only up to the second order; however, as long as the physical gate error rate $p_{\text{1-qubit}}$ and $p_{\text{2-qubit}}$ is at the same orders of magnitude as $p_{\text{phy }\ket{T}}$, $p_0$ is roughly at the same level as $p_{\text{phy }\ket{T}}$. In the end, we have the following:
\begin{equation}
    C_{\text{temporal}}(L)=(n-m_X)\sum\limits_{l=1}^{L}{\frac{1}{(1-p_l)^{n\left(\frac{n}{k}\right)^{L-l}}}}.
\end{equation}
Due to the fact that $p_0\ll1$, the time cost is dominated by the term corresponding to the first layer.
\begin{equation}
    C_{\text{temporal}}(L)=(n-m_X)\left(\frac{1}{(1-p_0)^n}+O\left(\frac{1}{(1-Cp_0^{\frac{d+1}{2}})^{\frac{n^2}{k}}}\right)\right).
\end{equation}
Finally, the spacetime cost product for $L$ layers of distillations is expressed as
\begin{equation}
\begin{aligned}
   C&=C_{\text{spatial}}\times C_{\text{temporal}}\\
   &=\left(\frac{1.5(m_X+k)}{k}\right)^L (n-m_X)\sum\limits_{l=1}^{L}{\frac{1}{(1-p_l)^{n\left(\frac{n}{k}\right)^{L-l}}}}d^3.
\end{aligned}
\end{equation}
In addition to the exact scalings of the cost, we also include some numerical examples estimated by Ref. \cite{magic_state_distillation_cost2} in the same table.
\begin{figure*}[t!]
    \centering
    \includegraphics[width=.8\linewidth]{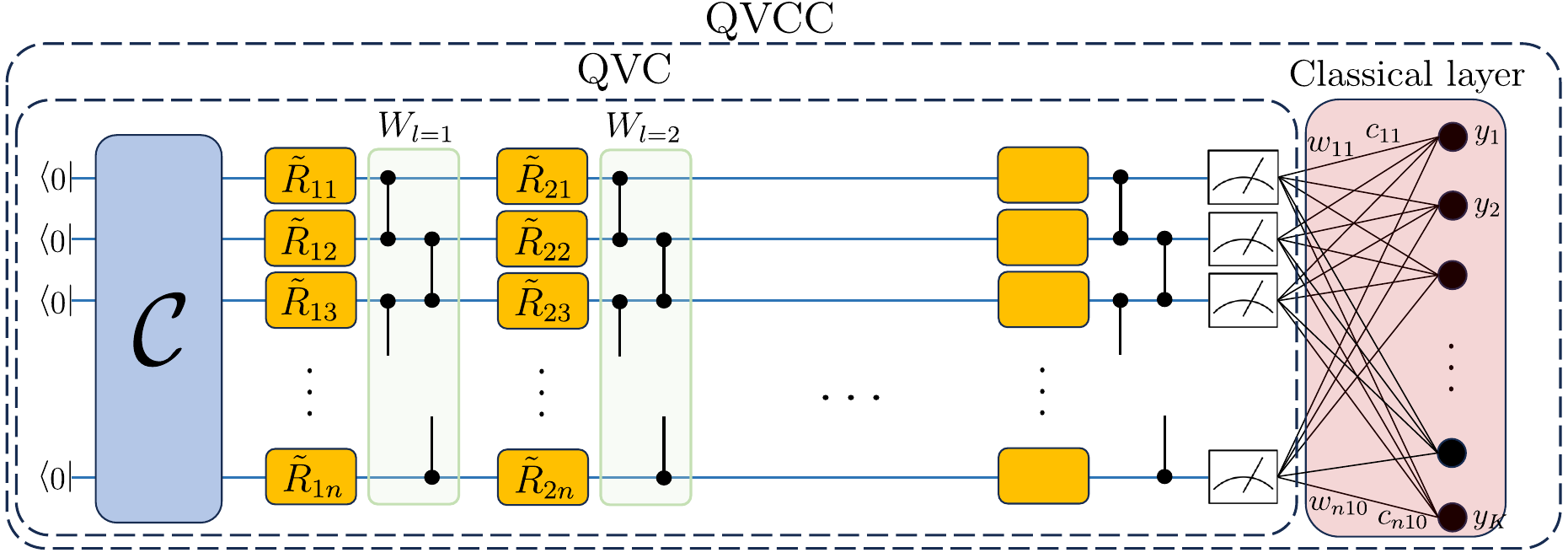}
    \caption{The circuit design of QVCC, where an additional fully connected layer is added after the variational circuit. The layout of the QVC component is exactly the same as in \Cref{fig: VQC}. $w_{ij},c_{ij}$ represent the weighting and shifting parameters in the layer, $y_i$ is the output.}
    \label{fig: QVCC}
\end{figure*}
\section{The design of QVCC}
As discussed in \Cref{subsec: depol channel}, the detailed circuit design of QVC with an additional fully connected layer, QVCC, is illustrated in \Cref{fig: QVCC}.

\section{Supplementary plots for depolarizing channel simulations}\label{appendix: supplementary plots for depol channel}
\begin{figure*}[tbhp!]
    \centering
    \includegraphics[width=\linewidth]{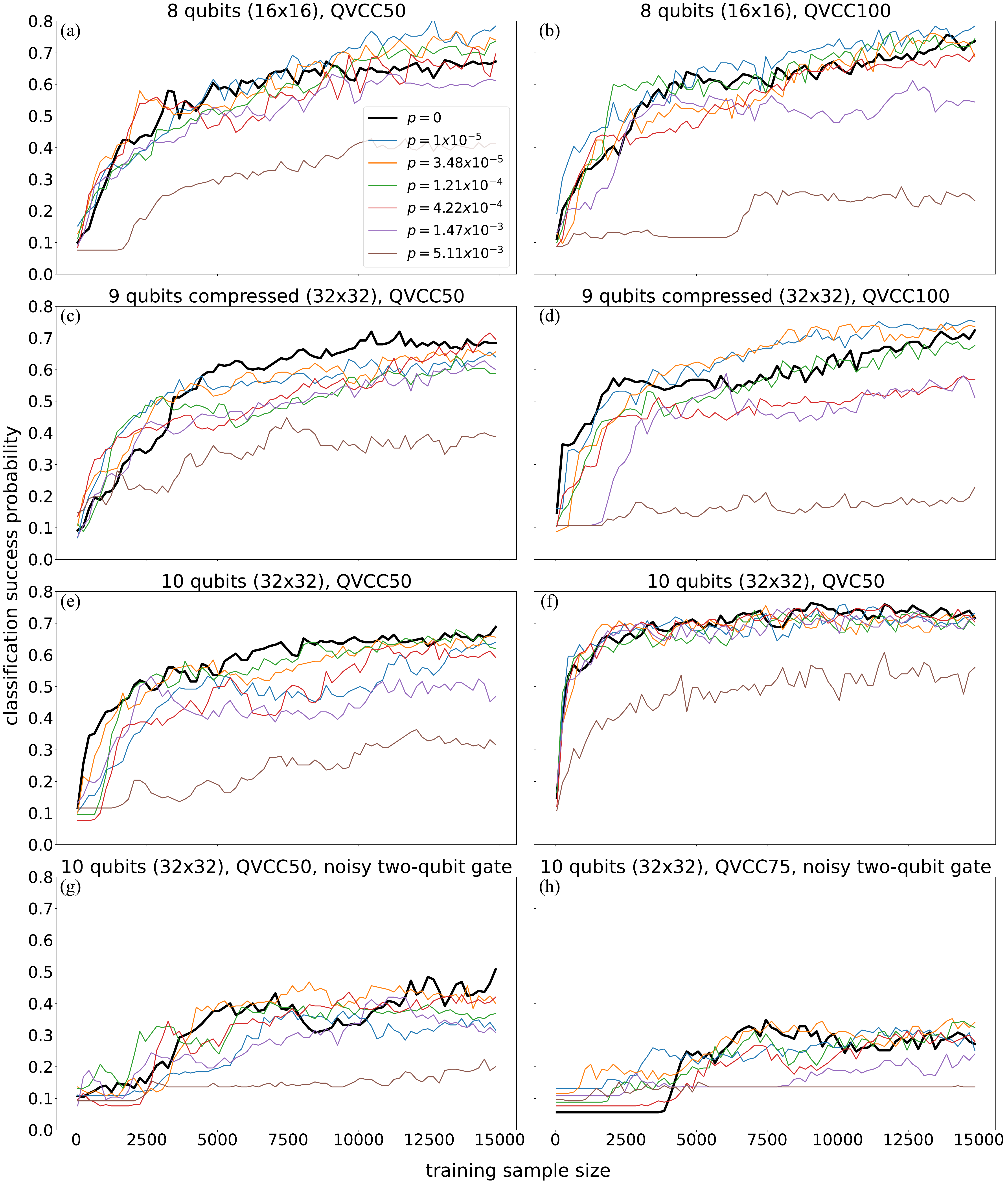}
    \caption{(a)--(h) Supplementary plots for the classification success rates range from 50 to 100 variational circuit layers, mostly on the QVCC model, and the qubit count ranges from 8 to 10. The structure of the plots is the same as in \Cref{fig: depol plots long layers}. For $9$-qubit case, a further compressed amplitude embedding \cite{amplitude_compressed_encoding} that encodes the pixel values onto both the real and imaginary parts of the amplitudes is used, which allows us to encode $2^{n+1}$ pixels for $n$ qubits.}
    \label{fig: classification_success_rate_short_layers}
\end{figure*}

\begin{figure*}[tbhp!]
    \centering
    \includegraphics[width=1\linewidth]{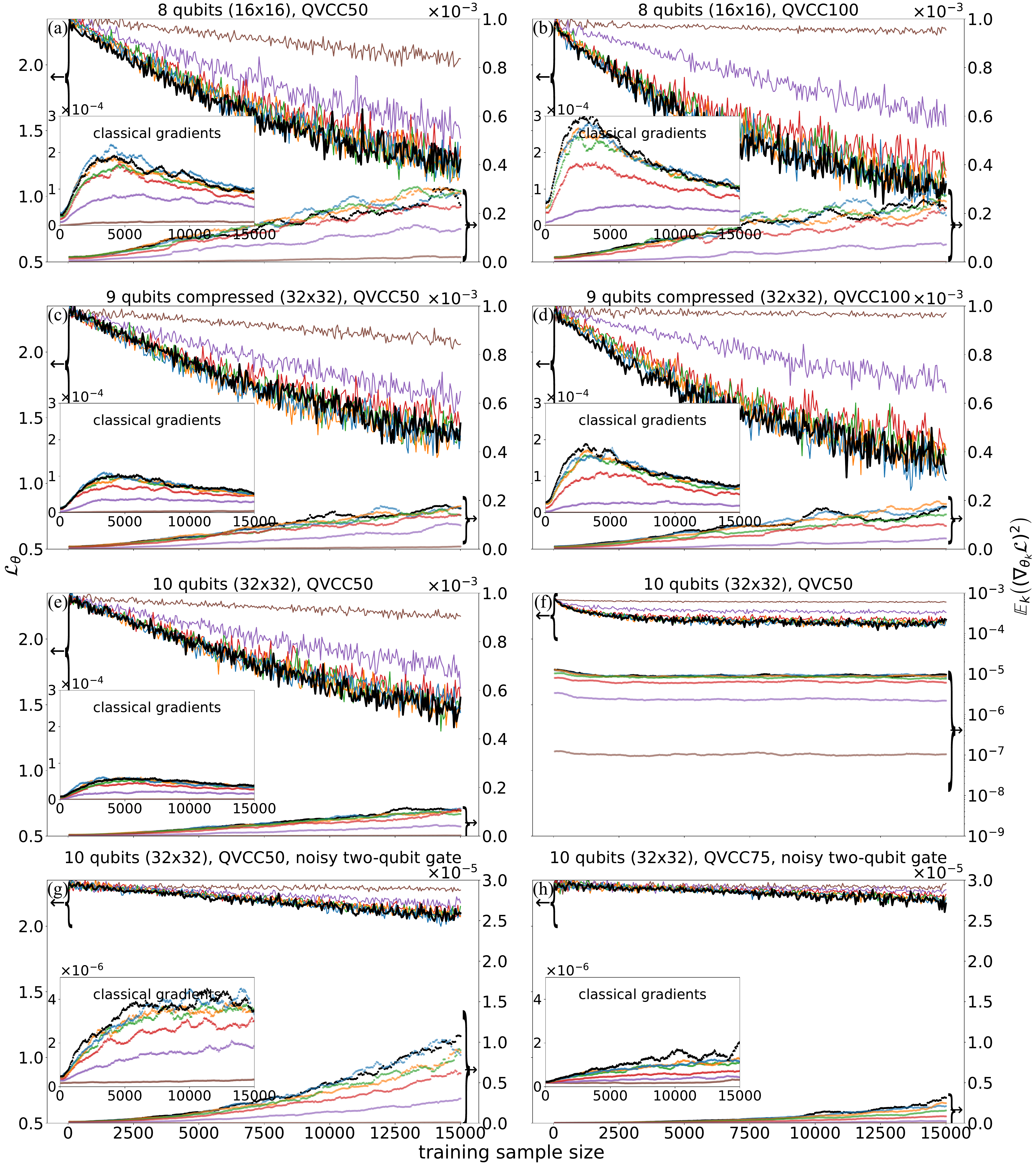}
    \caption{(a)--(h) Supplementary plots of the loss function and its average gradients squared from 50 to 100 variational circuit layers and 8 to 10 qubits. The structure and legends are the same as in \Cref{fig: depol plots long layers}. Note that to better demonstrate their behavior, the range of the right vertical axis (average squared gradients) for noisy two-qubit gate plots has been zoomed in, whereas the scaling for the loss function values is consistent throughout the eight plots.}
    \label{fig: loss_and_gradients_short_layers}
\end{figure*}

\begin{figure}
    \centering
    \includegraphics[width=1\linewidth]{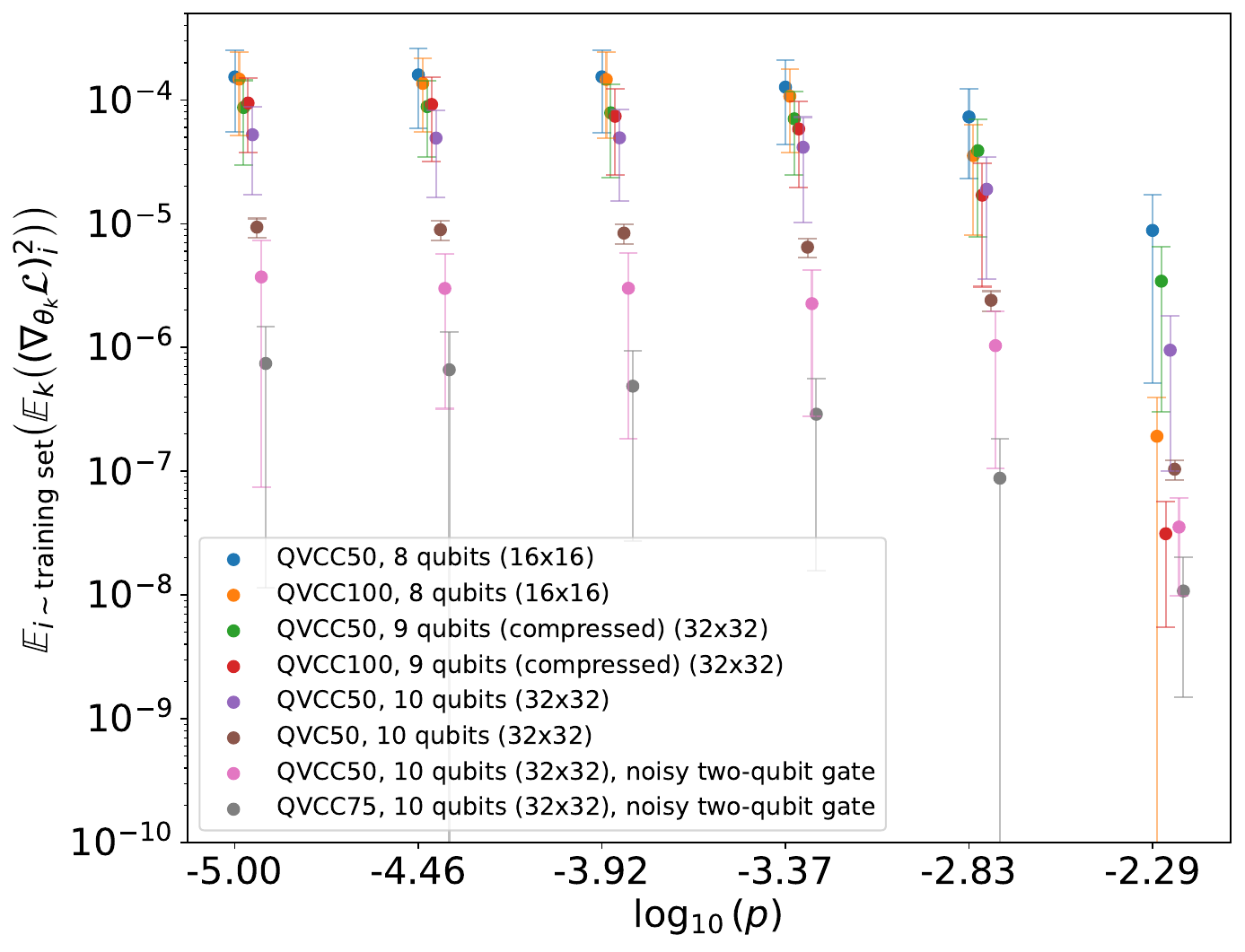}
    \caption{The corresponding squared gradients averaged over all parameters and training iterations versus the depolarizing strength $p$ for the simulations in \Cref{fig: loss_and_gradients_short_layers}. To prevent overlapping data points and error bars, results for different legends but the same depolarizing strength as marked by the tick labels on the horizontal axis, are slightly staggered on the plot. Apparently, the gradient decreases exponentially with respect to the depolarizing channel strength, the number of qubits, and the circuit depth.}
    \label{fig: gradients_vs_p_short_layers}
\end{figure}

We include some supplementary plots regarding the variational circuit simulations in \Cref{fig: classification_success_rate_short_layers,fig: loss_and_gradients_short_layers,fig: gradients_vs_p_short_layers} for the depolarizing channel on single-qubit gates, but with different numbers of layers, qubit counts, and whether a classical layer is included. In general, the behavior of the model during training is the same as that discussed in \Cref{subsec: depol channel}. However, due to a shorter (longer) circuit depth, the average size of the gradients is higher (lower) than observed in \Cref{fig: depol plots long layers,fig: gradients_vs_p_long_layers}, further underpinning the arguments about noise-induced barren plateaus \cite{noise_induced_barren_plateaus}. From \Cref{fig: loss_and_gradients_short_layers}, it is also clear that more qubits correspond to an exponentially smaller gradient as predicted by barren plateaus \cite{barren_plateaus}.

As shown in \Cref{fig: loss_and_gradients_short_layers}, in addition to the main observation discussed above, when a fully connected classical layer (QVCC model) is present at the end, we find that the cost function still has not fully converged after 15000 images trained. Simultaneously, the classification success rate is also not fully saturated, as illustrated in \Cref{fig: classification_success_rate_short_layers}. This may indicate the potential advantage of an extra classical layer in obtaining a higher success rate at saturation, provided that the variation of the observables $Z_i$ is still above the shot noise during the gradient descent. When the classical layer is applied, the gap in the trainability is even larger when contrasted with the noise-free two-qubit gates, as shown in \Cref{fig: classification_success_rate_short_layers,fig: loss_and_gradients_short_layers}.

Meanwhile, although the averaged gradients increase with more images trained, the zoomed figure of `classical gradients' in \Cref{fig: loss_and_gradients_short_layers}, which extracts the average gradients from the parameters of the classical layer only, reveals that these gradients still decay normally as the cost function approaches its minimum. The increasing gradients for the parameters in quantum layers arise due to increasing weights in the fully connected classical layer as more samples are trained, which also scales the gradients of quantum layers. Despite increasing gradients, we reiterate that barren plateaus are still present when conditions are met, as the Hilbert subspace explored by $\bm{\theta}$ remains unchanged, the Pauli-$Z$ observables remain concentrated, while the cost function and fluctuations due to shot noise scale simultaneously. 

\section{Depolarizing noise strength and logical $T$ gate error rate analysis}\label{appendix: Depolarising noise strength and actual gate error analysis}
Given a depolarizing channel with a depolarizing noise strength $p_{\text{depol}}$ added after an arbitrary single-qubit unitary $U$, the actual decay rate $p$ is determined from the standard randomized benchmarking \cite{randomised_benchmarking}. Specifically,
\begin{equation}\label{eq: randomized benchmarking fidelity}
    p\coloneqq \int_{}d\psi\text{Tr}(\ket{\psi}\bra{\psi} \Tilde{U}^{\dagger}\circ\Tilde{U}(\ket{\psi}\bra{\psi}))
\end{equation}
where $\Tilde{U}$ and $\Tilde{U}^{\dagger}$ are the channels of the noisy $U$ and $U^{\dagger}$ gates. Since the depolarizing channel commutes with any single-qubit unitaries, let $\ket{\psi}=V_{\psi}\ket{0}$, and absorb $V_{\psi}$ into $U$ as $W_{\psi}=UV_{\psi}$, we have
\begin{equation}\label{eq:randomised_benchmarking_error_rate_for_depolarising_channel}
\begin{aligned}
    &=\int d\psi\text{Tr}\{V_{\psi}\ket{0}\bra{0}V_{\psi}^{\dagger} U^{\dagger}\mathcal{E}^2_{\text{depol}}(UV_{\psi}\ket{0}\bra{0}V_{\psi}^{\dagger}U^{\dagger})U\}\\
    &=\int d\psi\text{Tr}\{W_{\psi}\ket{0}\bra{0}W_{\psi}^{\dagger}\mathcal{E}^2_{\text{depol}}(W_{\psi}\ket{0}\bra{0}W_{\psi}^{\dagger})\}\\
    &=\text{Tr}\{\ket{0}\bra{0}\mathcal{E}^2_{\text{depol}}(\ket{0}\bra{0})\}.
\end{aligned}
\end{equation}
Substituting $\mathcal{E}_{\text{depol}}\left(\frac{I+Z}{2}\right)=\frac{I+\left(1-\frac{4p_{\text{depol}}}{3}\right)Z}{2}$ into \Cref{eq:randomised_benchmarking_error_rate_for_depolarising_channel}, it is apparent that
\begin{equation}
    p=\frac{1+\left(1-\frac{4p_{\text{depol}}}{3}\right)^2}{2}.
\end{equation}
Therefore, the actual gate error rate is given by
\begin{equation}\label{eq: gate error to fidelity}
    r\coloneqq 1-p - (1-p)/d,
\end{equation}
where $d$ is the dimension of the Hilbert space, for a single-qubit state $d=2$. However, such an error rate for an arbitrary single-qubit unitary must be shared by a sequence of Clifford + $T$ gates in the practice of error correction code, and the number of $T$ gates required scales as $\sim1.5\log_2(1/\epsilon)$ for a precision of $\epsilon$ \cite{T_consumption_for_unitary}. Therefore, based on the assumption of partial QEC that only considers the significant error rate for $T$ gates, the relationship between the logical $T$ gate error rate $\epsilon_T$ to the arbitrary rotation is given as
\begin{equation}
    r\sim1-(1-\epsilon_T)^{1.5\log_2(1/\epsilon)}.
\end{equation}
In literature \cite{chamberland2020very,magic_state_distillation_cost1,magic_state_distillation_cost2}, a physical error rate of $10^{-4}$ was often used, which arises from the expectation that it has a reasonably good error suppression factor, which scales as $ O\left(\frac{p_{\text{threshold}}}{p}\right)$, and practical reachability. Since the partial QEC protocol would still require the physical gate error rate to be sufficiently below the threshold around $10^{-2}$ \cite{surface_codes}, at the time when full-scale QEC becomes feasible, a physical two-qubit error rate of $10^{-4}$ is fundamental and achievable, which is an order of magnitude lower than what has been achieved nowadays \cite{data_sheet}. According to the estimation in \cite{physical_T_and_logical_T_error_rate}, which uses the SI1000 error model \cite{SI1000} by assuming that $p_2=10p_1$, we have $\frac{2}{3}p_2\less\epsilon_T\less p_2$ for $p_2$ sufficiently small, where $p_1$, $p_2$ are the physical single and two-qubit gate error rates, respectively. Therefore, we use $\epsilon_T=10^{-4}$ as the upper bound for $p_2=10^{-4}$. For a logical $T$ gate with an raw error rate of $\epsilon_T=10^{-4}$, the final $U$ gate error rate can be derived as $r=1.99\cross 10^{-3}$ 
with a small tolerance of $\epsilon=10^{-4}$
in its rotation angle. This implies that the corresponding depolarizing channel strength is $p_{\text{depol}}=2.99\cross 10^{-3}$.

In one of the latest literature \cite{T_consumption_for_unitary_improved}, the authors improve the scaling of the number of $T$ gates required to be $\sim \log_2(1/\epsilon)$ by introducing $T_X$ gates that are $\pi/4$ rotations in $X$ direction, but have the same cost as normal $T$ gates. In this case, the actual gate error can be further relaxed to $1.33\cross 10^{-3}$, which corresponds to $p_{\text{depol}}=1.99\times10^{-3}$. At this rate, we show that the model is still trainable in \Cref{fig: depol plots long layers}, where the classification rate is only slightly decreased.

\section{Shot-noise estimation on the cross-entropy loss function}\label{appendix: variation in loss caused by shot noise}
For the variational circuits without any classical layer (QVC), the dependence of the loss function on the expectation values of the Pauli-$Z$ observables is found to be 
\begin{equation}
    \mathcal{L}_{\bm{\theta}}(\hat{\bm{z}}_i, \bm{y}_i)=\sum\limits_{j=1}^{n}{\bm{y}_{ij}\log\left(\text{softmax}\left(\hat{\bm{z}}_{ij}\right)\right)},
\end{equation}
where $\hat{\bm{y}}(\bm{x}_i)$ is replaced with $\hat{\bm{z}}_i$ which is simply the array of expectation values $\{\expval{Z_j}\}$ for input data with label $i$. We note that $\bm{y}_{ij}$, the true label that corresponds to the data, would only have one non-zero element for $j=a$, $\bm{y}_{ia}=1$, where $a$ is the correct label index. Therefore,
\begin{equation}
\begin{aligned}
    \mathcal{L}_{\bm{\theta}}(\hat{\bm{z}}_i, \bm{y}_i)&=\sum\limits_{j=1}^{n}{\bm{y}_{ij}\log\left(\text{softmax}\left(\hat{\bm{z}}_{ij}\right)\right)}\delta_{ja}\\
    &=\log\left(\text{softmax}\left(\hat{\bm{z}}_{ia}\right)\right).
\end{aligned}
\end{equation}
Taking the Taylor expansion, the variation of the loss function up to the first order is given by:
\begin{equation}
\begin{aligned}
    \delta\mathcal{L}_{\bm{\theta}}&=\sum\limits_{j}\frac{\partial\mathcal{L}_{\theta}}{\partial \hat{\bm{z}}_{ij}}\delta\hat{\bm{z}}_{ij}\\
    &=\frac{\partial\mathcal{L}_{\theta}}{\partial \hat{\bm{z}}_{ia}}\delta\hat{\bm{z}}_{ia}+\sum\limits_{j\ne a}\frac{\partial\mathcal{L}_{\theta}}{\partial \hat{\bm{z}}_{ij}}\delta\hat{\bm{z}}_{ij},
\end{aligned}
\end{equation}
where the term for $\hat{\bm{z}}_{ia}$ is purposely separated, which becomes apparent later.

We are interested in the limit where the shot noise becomes significant compared to the actual values of the gradient. In other words, when the model is not trainable, barren plateaus occur, and the variation in the loss function due to shot noise fluctuations is at the same level as the loss function gradients. In the limit of barren plateaus, the expectation values of the observables are concentrated around the equator of the Bloch sphere, which implies $\hat{\bm{z}}_{ij}\sim0$. After substituting the \texttt{softmax} function previously defined as in \Cref{eq: softmax definition}, and $\hat{\bm{z}}_{ij}\sim0$, the variation in the loss function is simplified to 
\begin{equation}
\begin{aligned}    
    \delta\mathcal{L}_{\bm{\theta}}&=\left(1-\frac{\exp(\hat{\bm{z}}_{ia})}{\sum\limits_{j}\exp(\hat{\bm{z}}_{ij})}\right)\delta \hat{\bm{z}}_{ia} - \sum\limits_{j\ne a}\frac{\exp(\hat{\bm{z}}_{ij})}{\sum\limits_{j}\exp(\hat{\bm{z}}_{ij})}\delta \hat{\bm{z}}_{ij}\\
    &=\frac{9}{10}\delta \hat{\bm{z}}_{ia}-\frac{1}{10}\sum\limits_{j\ne a}\delta \hat{\bm{z}}_{ij},
\end{aligned}
\end{equation}
where the partial derivative of the $\hat{\bm{z}}_{ia}$ term contains an extra 1. Further substituting $\delta\hat{\bm{z}}_{ij}=\sqrt{\frac{4p(1-p)}{N}}$ according to the standard deviation formula of the binomial distribution with $-1$ and $+1$ as the lower and upper bounds, where $p$ is the probability of measuring 0 in a single shot, and $N$ is the number of shots, it arrives at 
\begin{align}
    \delta\mathcal{L}_{\bm{\theta}}
    &=\frac{9}{10}\sqrt{\frac{4p(1-p)}{N}}-\frac{\sqrt{9}}{10}\sqrt{\frac{4p(1-p)}{N}}.
\end{align}
We underline that since $\delta\hat{\bm{z}}_{ij}$ is a random variable, it is rescaled by the square root upon summation. In the proximity of barren plateaus, the expectation values are concentrated at $\hat{\bm{z}}_{ij}\sim0$, it is easy to see $p\sim0.5$. Therefore, when $N=10000$, 
\begin{align}\label{eq:threshold for shot noise}
    \delta\mathcal{L}_{\bm{\theta}}\sim6\cross 10^{-3}.
\end{align}
We note that this estimation of the variation of the loss function is an upper bound since $\delta\hat{\bm{z}}_{ij}$ takes the standard deviation of the binomial distribution that is maximized for $\hat{\bm{z}}_{ij}\sim0$ and $p\sim0.5$. It guarantees the trainability of the model if the gradients are above such a value, and whether it is trainable below needs to be further discussed.

Contrasting this with the simulated results $E\left((\nabla_{\theta}\mathcal{L})^2\right)$ for the case without classical layer as in \Cref{fig: depol plots long layers,fig: loss_and_gradients_short_layers}, the point where the model starts to become distinctively non-trainable (as evident from the suppressed classification success rate in \Cref{fig: depol plots long layers,fig: classification_success_rate_short_layers}) is around $E\left((\nabla_{\theta}\mathcal{L})^2\right)\sim1.0\cross 10^{-7}$, which corresponds to gradients $3.16\cross 10^{-4}$. Clearly, this does not break the threshold in \Cref{eq:threshold for shot noise} where the model is trainable when the gradients are above $6\cross 10^{-3}$.

\renewcommand{\href}[2]{#2}
\bibliography{main}
\end{document}